\input harvmac
\overfullrule=0pt
\parindent 25pt
\tolerance=10000
%\draftmode
\input epsf

\newcount\figno
\figno=0
\def\fig#1#2#3{
\par\begingroup\parindent=0pt\leftskip=1cm\rightskip=1cm\parindent=0pt
\baselineskip=11pt
\global\advance\figno by 1
\midinsert
\epsfxsize=#3
\centerline{\epsfbox{#2}}
\vskip 12pt
{\bf Fig.\ \the\figno: } #1\par
\endinsert\endgroup\par
}
\def\figlabel#1{\xdef#1{\the\figno}}
\def\encadremath#1{\vbox{\hrule\hbox{\vrule\kern8pt\vbox{\kern8pt
\hbox{$\displaystyle #1$}\kern8pt}
\kern8pt\vrule}\hrule}}

\font\cmss=cmss10
\font\cmsss=cmss10 at 7pt

\def\inbar{\vrule height1.5ex width.4pt depth0pt}

 \def\frac#1#2{{#1\over #2}}

 \def\s{\sqrt}

 \def\al{\alpha'}
 \def\de{\partial}

 \def\f {\frac}
 \def\ti{\tilde}
 \def\ap{\alpha}

 \def\ddd{\cdot\cdot\cdot}
 
 \def\la{\langle}
 \def\lb{\rangle}

\def\IC{{\relax\,\hbox{$\inbar\kern-.3em{\rm C}$}}}
\def\IZ{\relax\ifmmode\mathchoice
{\hbox{\cmss Z\kern-.4em Z}}{\hbox{\cmss Z\kern-.4em Z}}
{\lower.9pt\hbox{\cmsss Z\kern-.4em Z}} {\lower1.2pt\hbox{\cmsss
Z\kern-.4em Z}}\else{\cmss Z\kern-.4em Z}\fi}
\def\IR{\relax{\rm I\kern-.18em R}}

\lref\Ma{
  J.~M.~Maldacena,
  ``The large N limit of superconformal field theories and supergravity,''
  Adv.\ Theor.\ Math.\ Phys.\  {\bf 2}, 231 (1998)
  [Int.\ J.\ Theor.\ Phys.\  {\bf 38}, 1113 (1999)]
  [arXiv:hep-th/9711200].
  %%CITATION = HEP-TH 9711200;%%
}

\lref\Wi{
  E.~Witten,
   ``Anti-de Sitter space, thermal phase transition, and confinement in  gauge
  theories,''
  Adv.\ Theor.\ Math.\ Phys.\  {\bf 2}, 505 (1998)
  [arXiv:hep-th/9803131].
  %%CITATION = HEP-TH 9803131;%%
}

\lref\Gu{
  S.~S.~Gubser,
  ``Thermodynamics of spinning D3-branes,''
  Nucl.\ Phys.\ B {\bf 551}, 667 (1999)
  [arXiv:hep-th/9810225].
  %%CITATION = HEP-TH 9810225;%%
}

\lref\HoMy{
  G.~T.~Horowitz and R.~C.~Myers,
   ``The AdS/CFT correspondence and a new positive energy conjecture for
  general relativity,''
  Phys.\ Rev.\ D {\bf 59}, 026005 (1999)
  [arXiv:hep-th/9808079].
  %%CITATION = HEP-TH 9808079;%%
}

\lref\My{
  R.~C.~Myers,
  ``Stress tensors and Casimir energies in the AdS/CFT correspondence,''
  Phys.\ Rev.\ D {\bf 60}, 046002 (1999)
  [arXiv:hep-th/9903203].
  %%CITATION = HEP-TH 9903203;%%
}

\lref\Ru{
  J.~G.~Russo,
  ``New compactifications of supergravities and large N {QCD},''
  Nucl.\ Phys.\ B {\bf 543}, 183 (1999)
  [arXiv:hep-th/9808117].
  %%CITATION = HEP-TH 9808117;%%
}

\lref\Ka{
  D.~Kabat,
  ``Black hole entropy and entropy of entanglement,''
  Nucl.\ Phys.\ B {\bf 453}, 281 (1995)
  [arXiv:hep-th/9503016].
  %%CITATION = HEP-TH 9503016;%%
}

\lref\HoSi{
  G.~T.~Horowitz and E.~Silverstein,
  ``The inside story: Quasilocal tachyons and black holes,''
  Phys.\ Rev.\ D {\bf 73}, 064016 (2006)
  [arXiv:hep-th/0601032].
  %%CITATION = HEP-TH 0601032;%%
}

\lref\RyTa{
  S.~Ryu and T.~Takayanagi,
  ``Holographic derivation of entanglement entropy from AdS/CFT,''
  Phys.\ Rev.\ Lett.\  {\bf 96}, 181602 (2006)
  [arXiv:hep-th/0603001];
  %%CITATION = HEP-TH 0603001;%%
 S.~Ryu and T.~Takayanagi,
  ``Aspects of holographic entanglement entropy,''
  JHEP {\bf 0608}, 045 (2006)
  [arXiv:hep-th/0605073].
  %%CITATION = HEP-TH 0605073;%%
}

\lref\HirTa{
  T.~Hirata and T.~Takayanagi,
  ``AdS/CFT and strong subadditivity of entanglement entropy,''
  arXiv:hep-th/0608213.
  %%CITATION = HEP-TH 0608213;%%
}

%%%%

\lref\Cardy{
P.~Calabrese and J.~L.~Cardy,
  ``Entanglement entropy and quantum field theory,''
  J.\ Stat.\ Mech.\  {\bf 0406}, P002 (2004)
  [arXiv:hep-th/0405152].
  %%CITATION = HEP-TH 0405152;%%
}

\lref\RuTe{
 J.~G.~Russo and A.~A.~Tseytlin,
  ``Magnetic flux tube models in superstring theory,''
  Nucl.\ Phys.\ B {\bf 461}, 131 (1996)
  [arXiv:hep-th/9508068].
  %%CITATION = HEP-TH 9508068;%%
}

\lref\TaUe{
 T.~Takayanagi and T.~Uesugi,
  ``Orbifolds as Melvin geometry,''
  JHEP {\bf 0112}, 004 (2001)
  [arXiv:hep-th/0110099].
  %%CITATION = HEP-TH 0110099;%%
}

\lref\KuMaMo{
  D.~Kutasov, J.~Marklof and G.~W.~Moore,
  ``Melvin models and diophantine approximation,''
  Commun.\ Math.\ Phys.\  {\bf 256}, 491 (2005)
  [arXiv:hep-th/0407150].
  %%CITATION = HEP-TH 0407150;%%
}

\lref\DuMo{E.~Dudas and J.~Mourad,
  ``D-branes in string theory Melvin backgrounds,''
  Nucl.\ Phys.\ B {\bf 622}, 46 (2002)
  [arXiv:hep-th/0110186].
  %%CITATION = HEP-TH 0110186;%%
}

\lref\TaUeD{
 T.~Takayanagi and T.~Uesugi,
  ``D-branes in Melvin background,''
  JHEP {\bf 0111}, 036 (2001)
  [arXiv:hep-th/0110200];
  %%CITATION = HEP-TH 0110200;%%
 T.~Takayanagi and T.~Uesugi,
  ``Flux stabilization of D-branes in NSNS Melvin background,''
  Phys.\ Lett.\ B {\bf 528}, 156 (2002)
  [arXiv:hep-th/0112199].
  %%CITATION = HEP-TH 0112199;%%
}

\lref\LMS{ H.~Liu, G.~W.~Moore and N.~Seiberg,
  ``Strings in a time-dependent orbifold,''
  JHEP {\bf 0206}, 045 (2002)
  [arXiv:hep-th/0204168];
  %%CITATION = HEP-TH 0204168;%%
``Strings in time-dependent orbifolds,''
  JHEP {\bf 0210}, 031 (2002)
  [arXiv:hep-th/0206182].
  %%CITATION = HEP-TH 0206182;%%
}

\lref\YaZw{
  H.~Yang and B.~Zwiebach,
  ``Rolling closed string tachyons and the big crunch,''
  JHEP {\bf 0508}, 046 (2005)
  [arXiv:hep-th/0506076];
  %%CITATION = HEP-TH 0506076;%%
H.~Yang and B.~Zwiebach,
  ``A closed string tachyon vacuum?,''
  JHEP {\bf 0509}, 054 (2005)
  [arXiv:hep-th/0506077].
  %%CITATION = HEP-TH 0506077;%%
}

\lref\Ts{
A.~A.~Tseytlin,
  ``Sigma model approach to string theory effective actions with tachyons,''
  J.\ Math.\ Phys.\  {\bf 42}, 2854 (2001)
  [arXiv:hep-th/0011033].
  %%CITATION = HEP-TH 0011033;%%
}

\lref\Se{
  A.~Sen,
  ``Tachyon dynamics in open string theory,''
  Int.\ J.\ Mod.\ Phys.\ A {\bf 20}, 5513 (2005)
  [arXiv:hep-th/0410103].
  %%CITATION = HEP-TH 0410103;%%
}

%%%%2

\lref\MoYa{
  N.~Moeller and H.~Yang,
  ``The nonperturbative closed string tachyon vacuum to high level,''
  arXiv:hep-th/0609208.
  %%CITATION = HEP-TH 0609208;%%
}

\lref\FaZaZa{ V.~Fateev, A.~B.~Zamolodchikov and
A.~B.~Zamolodchikov,
   ``Boundary Liouville field theory. I: Boundary state and boundary  two-point
  function,''
  arXiv:hep-th/0001012.
  %%CITATION = HEP-TH 0001012;%%
}

\lref\HaHo{
  S.~W.~Hawking and G.~T.~Horowitz,
  ``The Gravitational Hamiltonian, action, entropy and surface terms,''
  Class.\ Quant.\ Grav.\  {\bf 13}, 1487 (1996)
  [arXiv:gr-qc/9501014].
  %%CITATION = GR-QC 9501014;%%
}

\lref\HaOb{
  T.~Harmark and N.~A.~Obers,
  ``General definition of gravitational tension,''
  JHEP {\bf 0405}, 043 (2004)
  [arXiv:hep-th/0403103].
  %%CITATION = HEP-TH 0403103;%%
}

\lref\MaReview{
  E.~J.~Martinec,
  ``Defects, decay, and dissipated states,''
  arXiv:hep-th/0210231.
  %%CITATION = HEP-TH 0210231;%%
}

\lref\HeMiTa{
  M.~Headrick, S.~Minwalla and T.~Takayanagi,
  ``Closed string tachyon condensation: An overview,''
  Class.\ Quant.\ Grav.\  {\bf 21}, S1539 (2004)
  [arXiv:hep-th/0405064].
  %%CITATION = HEP-TH 0405064;%%
}

\lref\GuHeMiSc{
  M.~Gutperle, M.~Headrick, S.~Minwalla and V.~Schomerus,
  ``Space-time energy decreases under world-sheet RG flow,''
  JHEP {\bf 0301}, 073 (2003)
  [arXiv:hep-th/0211063].
  %%CITATION = HEP-TH 0211063;%%
}

\lref\DaGuHeMi{
  J.~R.~David, M.~Gutperle, M.~Headrick and S.~Minwalla,
  ``Closed string tachyon condensation on twisted circles,''
  JHEP {\bf 0202}, 041 (2002)
  [arXiv:hep-th/0111212].
  %%CITATION = HEP-TH 0111212;%%
}

\lref\MiTa{
  S.~Minwalla and T.~Takayanagi,
  ``Evolution of D-branes under closed string tachyon condensation,''
  JHEP {\bf 0309}, 011 (2003)
  [arXiv:hep-th/0307248].
  %%CITATION = HEP-TH 0307248;%%
}

%%%3

\lref\AdPoSi{
  A.~Adams, J.~Polchinski and E.~Silverstein,
  ``Don't panic! Closed string tachyons in ALE space-times,''
  JHEP {\bf 0110}, 029 (2001)
  [arXiv:hep-th/0108075].
  %%CITATION = HEP-TH 0108075;%%
}

\lref\Cv{
  M.~Cvetic {\it et al.},
  ``Embedding AdS black holes in ten and eleven dimensions,''
  Nucl.\ Phys.\ B {\bf 558}, 96 (1999)
  [arXiv:hep-th/9903214].
  %%CITATION = HEP-TH 9903214;%%
}

\lref\HaObT{
  T.~Harmark and N.~A.~Obers,
  ``Thermodynamics of spinning branes and their dual field theories,''
  JHEP {\bf 0001}, 008 (2000)
  [arXiv:hep-th/9910036].
  %%CITATION = HEP-TH 9910036;%%
}

\lref\CaHu{
  H.~Casini and M.~Huerta,
  ``A finite entanglement entropy and the c-theorem,''
  Phys.\ Lett.\ B {\bf 600}, 142 (2004)
  [arXiv:hep-th/0405111];
  %%CITATION = HEP-TH 0405111;%%
``A c-theorem for the entanglement entropy,''
  arXiv:cond-mat/0610375.
  %%CITATION = COND-MAT 0610375;%%
}

\lref\Hi{
  Y.~Hikida,
  ``Phase transitions of large N orbifold gauge theories,''
  arXiv:hep-th/0610119.
  %%CITATION = HEP-TH 0610119;%%
}

\lref\StTa{
  A.~Strominger and T.~Takayanagi,
  ``Correlators in timelike bulk Liouville theory,''
  Adv.\ Theor.\ Math.\ Phys.\  {\bf 7}, 369 (2003)
  [arXiv:hep-th/0303221].
  %%CITATION = HEP-TH 0303221;%%
}

\lref\HiTa{
  Y.~Hikida and T.~Takayanagi,
  ``On solvable time-dependent model and rolling closed string tachyon,''
  Phys.\ Rev.\ D {\bf 70}, 126013 (2004)
  [arXiv:hep-th/0408124].
  %%CITATION = HEP-TH 0408124;%%
}

\lref\Sc{
  V.~Schomerus,
  ``Rolling tachyons from Liouville theory,''
  JHEP {\bf 0311}, 043 (2003)
  [arXiv:hep-th/0306026].
  %%CITATION = HEP-TH 0306026;%%
}

\lref\McSi{
  J.~McGreevy and E.~Silverstein,
  ``The tachyon at the end of the universe,''
  JHEP {\bf 0508}, 090 (2005)
  [arXiv:hep-th/0506130];
  %%CITATION = HEP-TH 0506130;%%
 E.~Silverstein,
  ``Singularities and closed string tachyons,''
  arXiv:hep-th/0602230.
  %%CITATION = HEP-TH 0602230;%%
}

\lref\ALMSS{
  A.~Adams, X.~Liu, J.~McGreevy, A.~Saltman and E.~Silverstein,
  ``Things fall apart: Topology change from winding tachyons,''
  JHEP {\bf 0510}, 033 (2005)
  [arXiv:hep-th/0502021].
  %%CITATION = HEP-TH 0502021;%%
}

%%%4

\lref\Ho{
  G.~T.~Horowitz,
  ``Tachyon condensation and black strings,''
  JHEP {\bf 0508}, 091 (2005)
  [arXiv:hep-th/0506166].
  %%CITATION = HEP-TH 0506166;%%
}

\lref\NaReSu{
  Y.~Nakayama, S.~J.~Rey and Y.~Sugawara,
  ``The nothing at the beginning of the universe made precise,''
  arXiv:hep-th/0606127.
  %%CITATION = HEP-TH 0606127;%%
}

\lref\HiTai{
  Y.~Hikida and T.~S.~Tai,
  ``D-instantons and closed string tachyons in Misner space,''
  JHEP {\bf 0601}, 054 (2006)
  [arXiv:hep-th/0510129].
  %%CITATION = HEP-TH 0510129;%%
}

\lref\KaSt{
  J.~L.~Karczmarek and A.~Strominger,
  ``Matrix cosmology,''
  JHEP {\bf 0404}, 055 (2004)
  [arXiv:hep-th/0309138];
  %%CITATION = HEP-TH 0309138;%%
  ``Closed string tachyon condensation at c = 1,''
  JHEP {\bf 0405}, 062 (2004)
  [arXiv:hep-th/0403169].
  %%CITATION = HEP-TH 0403169;%%
}

\lref\DDLM{
  S.~R.~Das, J.~L.~Davis, F.~Larsen and P.~Mukhopadhyay,
  ``Particle production in matrix cosmology,''
  Phys.\ Rev.\ D {\bf 70}, 044017 (2004)
  [arXiv:hep-th/0403275].
  %%CITATION = HEP-TH 0403275;%%
}

\lref\TaLD{
  T.~Takayanagi,
  ``Matrix model and time-like linear dilaton matter,''
  JHEP {\bf 0412}, 071 (2004)
  [arXiv:hep-th/0411019].
  %%CITATION = HEP-TH 0411019;%%
}

\lref\RTF{
  J.~G.~Russo and A.~A.~Tseytlin,
  ``Supersymmetric fluxbrane intersections and closed string tachyons,''
  JHEP {\bf 0111}, 065 (2001)
  [arXiv:hep-th/0110107].
  %%CITATION = HEP-TH 0110107;%%
}

\lref\Po{
  J.~Polchinski,
  ``String theory. Vol. 1: An introduction to the bosonic string,''
  {\it Cambridge, UK: Univ. Pr.} (1998) 402 p.
%\href{http://www.slac.stanford.edu/spires/find/hep/www?irn=4634799}{SPIRES entry}
}

\lref\BKLS{
  L.~Bombelli, R.~K.~Koul, J.~H.~Lee and R.~D.~Sorkin,
  ``A quantum source of entropy for black holes,''
  Phys.\ Rev.\ D {\bf 34}, 373 (1986).
  %%CITATION = PHRVA,D34,373;%%
}

\lref\Sr{
  M.~Srednicki,
  ``Entropy and area,''
  Phys.\ Rev.\ Lett.\  {\bf 71}, 666 (1993)
  [arXiv:hep-th/9303048].
  %%CITATION = HEP-TH 9303048;%%
}

%%%%5

\lref\FFN{
  P.~Fendley, M.~P.~A.~Fisher and C.~Nayak,
  ``Topological Entanglement Entropy from the Holographic Partition Function,''
  arXiv:cond-mat/0609072.
  %%CITATION = COND-MAT 0609072;%%
}

\lref\CoGu{
 M.~S.~Costa and M.~Gutperle,
  ``The Kaluza-Klein Melvin solution in M-theory,''
  JHEP {\bf 0103}, 027 (2001)
  [arXiv:hep-th/0012072];
  %%CITATION = HEP-TH 0012072;%%
  M.~Gutperle,
   ``A note on perturbative and nonperturbative instabilities of twisted
  circles,''
  Phys.\ Lett.\ B {\bf 545}, 379 (2002)
  [arXiv:hep-th/0207131].
  %%CITATION = HEP-TH 0207131;%%
}

\lref\ClMa{
  R.~Clarkson and R.~B.~Mann,
  ``Eguchi-Hanson solitons in odd dimensions,''
  Class.\ Quant.\ Grav.\  {\bf 23}, 1507 (2006)
  [arXiv:hep-th/0508200].
  %%CITATION = HEP-TH 0508200;%%
}

\lref\BiDeMu{
  A.~Biswas, T.~K.~Dey and S.~Mukherji,
  ``R-charged AdS bubble,''
  Phys.\ Lett.\ B {\bf 613}, 208 (2005)
  [arXiv:hep-th/0412124].
  %%CITATION = HEP-TH 0412124;%%
}

\lref\BaRo{
  V.~Balasubramanian and S.~F.~Ross,
  ``The dual of nothing,''
  Phys.\ Rev.\ D {\bf 66}, 086002 (2002)
  [arXiv:hep-th/0205290];
  %%CITATION = HEP-TH 0205290;%%
  D.~Birmingham and M.~Rinaldi,
  ``Bubbles in anti-de Sitter space,''
  Phys.\ Lett.\ B {\bf 544}, 316 (2002)
  [arXiv:hep-th/0205246].
  %%CITATION = HEP-TH 0205246;%%
}

\lref\BeHi{
  O.~Bergman and S.~Hirano,
  ``Semi-localized instability of the Kaluza-Klein linear dilaton vacuum,''
  Nucl.\ Phys.\ B {\bf 744}, 136 (2006)
  [arXiv:hep-th/0510076].
  %%CITATION = HEP-TH 0510076;%%
}

\lref\HLW{C.~Holzhey, F.~Larsen and F.~Wilczek,
  ``Geometric and renormalized entropy in conformal field theory,''
  Nucl.\ Phys.\ B {\bf 424}, 443 (1994)
  [arXiv:hep-th/9403108].
  %%CITATION = HEP-TH 9403108;%%
}

\lref\Fu{
  D.~V.~Fursaev,
  ``Proof of the holographic formula for entanglement entropy,''
  JHEP {\bf 0609}, 018 (2006)
  [arXiv:hep-th/0606184].
  %%CITATION = HEP-TH 0606184;%%
}

\lref\CaHuM{
H.~Casini and M.~Huerta,
   ``Entanglement and alpha entropies for a massive scalar field in two
  dimensions,''
  J.\ Stat.\ Mech.\  {\bf 0512}, P012 (2005)
  [arXiv:cond-mat/0511014].
  %%CITATION = COND-MAT 0511014;%%
}

\lref\SuUg{
  L.~Susskind and J.~Uglum,
  ``Black hole entropy in canonical quantum gravity and superstring theory,''
  Phys.\ Rev.\ D {\bf 50}, 2700 (1994)
  [arXiv:hep-th/9401070].
  %%CITATION = HEP-TH 9401070;%%
}

%%6

\lref\SuMe{
  T.~Suyama,
  ``Properties of string theory on Kaluza-Klein Melvin background,''
  JHEP {\bf 0207}, 015 (2002)
  [arXiv:hep-th/0110077].
  %%CITATION = HEP-TH 0110077;%%
}

\lref\SuRG{
  T.~Suyama,
  ``Closed string tachyons and RG flows,''
  JHEP {\bf 0210}, 051 (2002)
  [arXiv:hep-th/0210054].
  %%CITATION = HEP-TH 0210054;%%
}

\lref\MaMo{
  E.~J.~Martinec and G.~W.~Moore,
  ``On decay of K-theory,''
  arXiv:hep-th/0212059.
  %%CITATION = HEP-TH 0212059;%%
}

\lref\CSV{
  B.~Craps, S.~Sethi and E.~P.~Verlinde,
  ``A matrix big bang,''
  JHEP {\bf 0510}, 005 (2005)
  [arXiv:hep-th/0506180].
  %%CITATION = HEP-TH 0506180;%%
}

\lref\Ha{ A.~Hashimoto and K.~Thomas, ``Dualities, twists, and gauge
theories with non-constant non-commutativity,'' JHEP {\bf 0501}, 033
(2005) [arXiv:hep-th/0410123];
  %%CITATION = HEP-TH 0410123;%%
  ``Non-commutative gauge theory on D-branes in Melvin universes,''
  JHEP {\bf 0601}, 083 (2006)
  [arXiv:hep-th/0511197].
  %%CITATION = HEP-TH 0511197;%%
}

\lref\AMMW{
  O.~Aharony, J.~Marsano, S.~Minwalla and T.~Wiseman,
   ``Black hole - black string phase transitions in thermal 1+1 dimensional
  supersymmetric Yang-Mills theory on a circle,''
  Class.\ Quant.\ Grav.\  {\bf 21}, 5169 (2004)
  [arXiv:hep-th/0406210].
  %%CITATION = HEP-TH 0406210;%%
}

\lref\MiYi{
  Y.~Michishita and P.~Yi,
  ``D-brane probe and closed string tachyons,''
  Phys.\ Rev.\ D {\bf 65}, 086006 (2002)
  [arXiv:hep-th/0111199].
  %%CITATION = HEP-TH 0111199;%%
}

\lref\BiDa{
 N.~D.~Birrell and P.~C.~W.~Davies,
  ``Quantum Fields In Curved Space,''
%\href{http://www.slac.stanford.edu/spires/find/hep/www?irn=998621}{SPIRES entry}
{\it Cambridge, Uk: Univ. Pr.} ( 1982) 340p.
}

\lref\HoKa{
  K.~Hori and A.~Kapustin,
   ``Duality of the fermionic 2d black hole and N = 2 Liouville theory as
  mirror symmetry,''
  JHEP {\bf 0108}, 045 (2001)
  [arXiv:hep-th/0104202].
  %%CITATION = HEP-TH 0104202;%%
}

\lref\Ah{
  O.~Aharony, J.~Marsano, S.~Minwalla, K.~Papadodimas and M.~Van Raamsdonk,
   ``The Hagedorn / deconfinement phase transition in weakly coupled large N
  gauge theories,''
  Adv.\ Theor.\ Math.\ Phys.\  {\bf 8}, 603 (2004)
  [arXiv:hep-th/0310285].
  %%CITATION = HEP-TH 0310285;%%
}

\lref\GTT{
  S.~Gukov, T.~Takayanagi and N.~Toumbas,
  ``Flux backgrounds in 2D string theory,''
  JHEP {\bf 0403}, 017 (2004)
  [arXiv:hep-th/0312208].
  %%CITATION = HEP-TH 0312208;%%
}

\lref\Sud{
  B.~Sundborg,
  ``The Hagedorn transition, deconfinement and N = 4 SYM theory,''
  Nucl.\ Phys.\ B {\bf 573}, 349 (2000)
  [arXiv:hep-th/9908001].
  %%CITATION = HEP-TH 9908001;%%
}

\lref\She{
  J.~H.~She,
  ``A matrix model for Misner universe,''
  JHEP {\bf 0601}, 002 (2006)
  [arXiv:hep-th/0509067];
  %%CITATION = HEP-TH 0509067;%%
  ``Winding string condensation and noncommutative deformation of spacelike
  singularity,''
  Phys.\ Rev.\ D {\bf 74}, 046005 (2006)
  [arXiv:hep-th/0512299].
  %%CITATION = HEP-TH 0512299;%%
}

\lref\AJ{
  D.~Astefanesei and G.~C.~Jones,
  ``S-branes and (anti-)bubbles in (A)dS space,''
  JHEP {\bf 0506}, 037 (2005)
  [arXiv:hep-th/0502162].
  %%CITATION = HEP-TH 0502162;%%
}

\lref\SAM{
  D.~Astefanesei, R.~B.~Mann and C.~Stelea,
  ``Nuttier bubbles,''
  JHEP {\bf 0601}, 043 (2006)
  [arXiv:hep-th/0508162].
  %%CITATION = HEP-TH 0508162;%%
}

\baselineskip 18pt plus 2pt minus 2pt

\Title{\vbox{\baselineskip12pt \hbox{hep-th/0611035}
\hbox{KUNS-2050}
  }}
{\vbox{\centerline{AdS Bubbles, Entropy and Closed String Tachyons}
}}

\centerline{Tatsuma
Nishioka\foot{e-mail:nishioka@gauge.scphys.kyoto-u.ac.jp} and
Tadashi Takayanagi\foot{e-mail:takayana@gauge.scphys.kyoto-u.ac.jp}}
\medskip\centerline{Department of Physics, Kyoto University, Kyoto
606-8502, Japan}

\vskip .5in

\centerline{\bf Abstract} We study the conjectured connection
between AdS bubbles (AdS solitons) and closed string tachyon
condensations. We confirm that the entanglement entropy, which
measures the degree of freedom, decreases under the tachyon
condensation. The entropies in supergravity and free Yang-Mills
agree with each other remarkably. Next we consider the tachyon
condensation on the AdS twisted circle and argue that its endpoint
is given by the twisted AdS bubble, defined by the double Wick
rotation of rotating black 3-brane solutions. We calculated the
Casimir energy and entropy and checked the agreements between the
gauge and gravity results. Finally we show an infinite boost of a
null linear dilaton theory with a tachyon wall (or bubble), leads to
a solvable time-dependent background with a bulk tachyon
condensation. This is the simplest example of spacetimes with null
boundaries in string theory.

\noblackbox

\Date{November, 2006}

%\listtoc
\writetoc

\newsec{Introduction}

The understandings of closed string tachyon condensation have been
far from complete in spite of many efforts e.g. the pioneering work
by Adams, Polchinski and Silverstein \AdPoSi\ (a list of references
can be found in reviews \MaReview \HeMiTa). One of the most
important problems is to examine the time-dependent dynamical
process of tachyon condensation.

Recently, a remarkable progress has been made by Horowitz and
Silverstein \Ho \HoSi\ employing the AdS/CFT correspondence \Ma.
They considered an unstable configuration of the near horizon
geometry of D3-branes by putting the anti-periodic boundary
condition for fermions. There appears a closed string tachyon which
is localized in a finite region of the spacetime \ALMSS \Ho \BeHi
\HoSi. The endpoint of the tachyon condensation is conjectured to be
the static AdS bubble solution (or AdS soliton) \Ho \HoSi. This
process of tachyon condensation has an equivalent description in the
dual Yang-Mills theory. The dynamics of tachyon condensation is
mapped to the more traditional problem of time-dependent process in
strongly coupled gauge theories.

In this paper, we would like to explore this scenario further. First
we show that the degree of freedom decreases under this tachyon
condensation process by computing the entanglement entropy \BKLS \Sr
\RyTa\ of the dual Yang-Mills theory\foot{Notice that the
thermodynamical entropy is zero for this solution since we are
considering zero temperature.}. In general, the degree of freedom is
expected to decrease under the closed string tachyon condensation
since the radiations produced by the process will carry away a part
of it. We will also present the dual holographic
computation\foot{Quite recently, a slightly analogous holographic
relation about the entanglement entropy in 3D topological QFTs has
been pointed out in \FFN. There, the boundary entropy (or
$g-$function) is dual to the bulk topological entanglement entropy.}
\RyTa\ in supergravity. This analysis of the entanglement entropy
offers a new evidence for the conjectured scenario of the closed
string tachyon condensation, in addition to the known decreasing of
energy density \HoMy \HoSi.

Next we consider a near horizon geometry of D3-branes with a twisted
boundary condition. The dual geometry is described by the AdS
geometry with the twisted identification. It is equivalent to a
twisted circle or Melvin background (refer to e.g. the papers \RuTe
\CoGu \SuMe \TaUe \RTF \DuMo \TaUeD \MiYi \DaGuHeMi \HeMiTa \Ha )
fibred over the radial direction of the AdS. Its radius of the
circle shrinks toward the IR region. Then the closed string theory
in this background has a tachyon field localized both in the IR
region and in a certain $S^3$ inside the $S^5$. We will claim that
the end point of the tachyon condensation is given by the bubble
solution obtained from the double Wick rotation of the rotating
D3-brane solution. We will check this claim by computing the energy
density (Casimir energy) and entanglement entropy in both the free
Yang-Mills and gravity theory. We find qualitative agreements
between them in general. Remarkably, the entropy in the near
extremal region precisely agrees with each other including the
numerical factor. We may also think this as a further evidence for
AdS/CFT correspondence in a slightly non-BPS background. We also
observe the qualitative agreement between the ADM energy of the
twisted AdS bubble and the Casimir energy of the dual free
Yang-Mills.

Another way to study the dynamical process of tachyon condensation
is to directly construct the corresponding time-dependent
backgrounds of string theory. Recently, its possible relevance to a
resolution of cosmological singularities has been discussed in
\McSi. In general, it is very difficult to find a well-controllable
time-dependent model with closed string tachyon condensation.

A simple example of the bulk tachyon condensation in bosonic string
(or type 0 string) has been proposed to be described by the
time-like Liouville theory \StTa \Sc\ (see e.g. \KaSt \HiTa \TaLD
\McSi \She \HiTai \HoSi \NaReSu\ for a partial list of further
progresses). However, this theory has not been completely
understood, especially because the continuation from the Euclidean
theory is not straightforward; there is a potential ambiguity with
respect to the choice of vacua in a time-dependent background. In
the last part of the present paper, we give a simpler solvable model
which describes bulk closed string tachyon condensation. This is
obtained by the infinite boost of the null linear dilaton background
with a Liouville potential. Equivalently we can regard the
background as a flat spacetime with a null boundary. Since there has
been no general answer to what kinds of boundaries are allowed in a
spacetime of string theory, this offers an useful basic example.

After we completed computations in this paper, we noticed an
interesting paper \Hi\ which discusses the analogous scenario of the
closed string tachyon condensation via the AdS/CFT correspondence in
a different model \ClMa (see also \SAM).

The organization of this paper is as follows. In section 2, we first
review the conjectured connection between AdS bubbles and closed
string tachyon condensation. Then we provide a further evidence for
this conjecture by computing the entanglement entropy. In section 3,
we consider the twisted AdS bubble solution and claim that it is the
endpoint of the closed string tachyon condensation on D3-branes
wrapped on the twisted circle by computing the energy and entropy.
In section 4, we present a simple construction of a spacetime with a
null boundary via bulk closed string tachyon condensation. In
section 5 we summarize conclusions.

\newsec{AdS Bubbles and Closed String Tachyons}

\subsec{Static AdS Bubble Solution}

Consider $N$ D3-branes in type IIB string. The world volume
coordinates are denoted by $(t,\chi,x_1,x_2)$. We compactify $\chi$
with period $L$ and put the anti-periodic boundary condition for all
fermions. Its near horizon geometry is represented by the $AdS_5$
\eqn\adsdth{ds^2=R^2
\f{dr^2}{r^2}+\f{r^2}{R^2}(-dt^2+d\chi^2+dx_1^2+dx_2^2).} The
important point is that the radius of the thermal circle $\chi$ gets
smaller as we goes into the IR region $r\to 0$. Thus we expect that
when its radius $\f{rL}{R}$ is of order $l_s$ (i.e. string scale), a
closed string tachyon appears. This tachyon is clearly localized in
the IR region. To make this more precise, we can start with a shell
of D3-branes which is described by the ten dimensional metric of
type IIB supergravity.
\eqn\shell{ds^2=h^{-1}(r)[-dt^2+d\chi^2+dx_1^2+dx_2^2]+h(r)
(dr^2+r^2d\Omega_5^2),} where $h(r)$ is the function defined by
\eqn\deffin{h(r)=\f{R^2}{r^2}\ \ (r>r_0),\ \ \ h(r)=\f{R^2}{r_0^2} \
\ (r\leq r_0).} Then the inside of the shell ($r<r_0$) the metric is
flat and thus we can employ the familiar perturbative world-sheet
analysis on the existence of closed string tachyons.

The remarkable claim made by Horowitz and Silverstein \HoSi\ is that
this unstable background decays into the static bubble \Wi\
\eqn\bubble{ds^2=R^2\f{dr^2}{r^2f(r)}+\f{r^2}{R^2}
(-dt^2+f(r)d\chi^2+dx_1^2+dx_2^2),} where $f(r)=1-(r_0/r)^4$. This
can be obtained from the double wick rotation of the
AdS-Schwartzschild solution and is called the (static) AdS
bubble\foot{Refer to e.g. \BaRo \AJ\ for a time-dependent bubble
solution obtained by another double Wick rotation of the AdS black
hole.} or AdS soliton \HoMy. Near the point $r=r_0$ the metric is
approximated by $ds^2\simeq dy^2+\f{4r_0^2}{R^4}y^2d\chi^2$, where
$y^2\equiv R^2(r-r_0)/r_0$. Thus to make the metric regular at this
point the periodicity $L$ of $\chi$ should be
\eqn\peridochi{L=\f{\pi R^2}{r_0}.}

In the dual Yang-Mills theory, this closed string tachyon
condensation is interpreted as follows \HoSi. The near horizon limit
of the D3-brane shell \shell\ corresponds to the supersymmetric
vacuum of $N=4$ super Yang-Mills theory with non-zero expectation
values of transverse scalar fields. Now we compactify one of the
three space coordinates and put the anti-periodic boundary condition
for all fermions. Then the supersymmetry is completely broken and
the scalar fields acquire non-zero masses from radiative
corrections. Thus the Coulomb branch is lifted and the theory
becomes almost the same as the pure Yang-Mills, which shows the
confinement behavior \Wi. The cut off of the IR region $r>r_0$ in
the bubble solution \bubble\ corresponds to the mass gap due to this
confinement.

\subsec{Casimir Energy}

An important evidence for this conjecture is that the AdS soliton
has the lowest energy\foot{Refer to \GuHeMiSc\ for the proof that
the energy is decreasing under the closed string tachyon
condensation in asymptotically flat spaces.} \HoMy \HoSi\ given by
\eqn\energy{\f{E}{V_2}=-\f{\pi^3
R^3}{16G^{(5)}_NL^3}=-\f{\pi^2}{8}\cdot\f{N^2}{L^3},} where\foot{
Also we have employed the standard relation $G^{(5)}=\f{\pi
R^3}{2N^2}=\f{G^{(10)}}{\pi^3R^5}$ in the type IIB string on
$AdS_5\times S^5$.} $G^{(5)}_N$ is the 5D Newton constant, and $V_2$
is the infinite volume of $(x_1,x_2)$. Here we have used the
definition of the energy  in an asymptotically AdS space \HaHo \HaOb
\eqn\energydef{E=-\f{1}{8\pi G^{(5)}_N}\int_{S} N(K-K_{0}),} where
the integral is over a surface near infinity $S$. $N$ is defined
such that the norm of the time-like Killing field is $-N^{2}$. $K$
is the trace of the extrinsic curvature of this surface. $K_{0}$ is
the trace of the extrinsic curvature of a surface with the same
intrinsic geometry in the background spacetime. The energy in an
asymptotically AdS space is defined such that the AdS space itself
has the vanishing energy $E=0$ \HaHo \HoMy.

It is useful to compare the above energy with the Casimir energy
computed in the free Yang-Mills theory \HoMy \My. Consider a
massless real scalar field\foot{We normalized the field such that
the Lagrangian is given by $L=\f{1}{2}(\de_\mu\phi)^2$.}
$\phi(t,\chi,x_1,x_2)(=\phi(x))$. We compactify the $\chi$ direction
such that $\chi\sim \chi+L$. Then the two point function can be
found to be \eqn\correlation{\la \phi(x)\phi(x')\lb
=\f{1}{4\pi^2}\sum_{n\in {\bf
Z}}\f{1}{(x_1-x'_1)^2+(x_2-x'_2)^2+(\chi-\chi'-nL)^2-(t-t')^2}.} The
energy density
$T_{00}=\f{1}{2}\left[(\de_0\phi)^2+\sum_i(\de_i\phi)^2\right]$ can
be obtained by the point splitting regularization (refer to \BiDa
\My\ for details) \eqn\point{\lim_{x\to x'}
\f{1}{2}\left[\de_0\phi(x)\de'_0\phi(x')
+\de_i\phi(x)\de^{'}_{i}\phi(x')\right].}
This leads to \eqn\resultca{T_{00}=\f{1}{4\pi^2}\sum_{n\neq
0}\left(-\f{2}{(L n)^4}\right)=-\f{\pi^2}{90L^4},} where we
regularize the summation by excluding the divergent term $n=0$. We
can perform the similar analysis for a free Majorana fermion $\psi$
with the anti-periodic boundary condition $\psi(z+L)=-\psi(z)$ and
obtain the energy density for each component
\eqn\eenrgyd{T_{00}=\f{1}{4\pi^2}\sum_{n\neq 0}\left(\f{2(-1)^n}{(
nL)^4}\right)=-\f{7\pi^2}{720 L^4}.} In the free N=4 super
Yang-Mills, there are $8N^2$ bosons and $8N^2$ fermions and thus we
finally obtain\foot{Here we are implicitly using the fact that the
contribution of the gauge fields is the same as that of two real
scalar fields.} \eqn\nfouren{\f{E}{V_2L}=T_{00}=8N^2\cdot
\left(-\f{\pi^2}{90L^4}\right)+8N^2\cdot
\left(-\f{7\pi^2}{720L^4}\right)=-\f{\pi^2N^2}{6L^4}.} This agrees
with \energy\ up to the factor $4/3$ \HoMy. Since the gravity
description corresponds to the strongly coupled limit of the
Yang-Mills theory, we can say that this agreement is rather
excellent.

\subsec{Entanglement Entropy: Gravity Side}

Now we wish to turn to another quantity called the entanglement
entropy \BKLS \Sr \RyTa\ as another evidence for the closed string
tachyon condensation. Divide the space manifold (in our case it is
$R^2\times S^1$) into two parts $A$ and $B$, and trace out the
Hilbert space for the subsystem $B$. This procedure defines the
reduced density matrix $\rho_A$ for the subsystem $A$. Then the
entanglement entropy $S_A$ is defined by the von-Neumann entropy
$S_A=-\tr \rho_A\log \rho_A$ with respect to the reduced density
matrix $\rho_A$. This leads to a non-vanishing entropy even if we
start with a pure state on the total space $A\cup B$. The choice of
$A$ is arbitrary and we can define infinitely many entropies $S_A$
accordingly.

In general, the entanglement entropy measures the degree of freedom
and thus we would like to claim that {\it the entanglement entropy
in the dual Yang-Mills theory should decrease under the closed
string tachyon condensation}. Indeed, in two dimension the entropy
is essentially known to be proportional to the central charge $c$
\HLW \Cardy. However, we should keep in mind that the UV behavior of
$S_A$ will not change under the localized closed string tachyon
condensation. Thus the divergent piece of the entropy, which is
proportional to the area of the boundary $\de A$ of the subsystem
$A$ (known as the area law \BKLS \Sr), will not change because this
part is only sensitive to UV quantities. We expect that only a
finite part of the entropy will change. Thus we will consider the
difference between the entropy before and after the tachyon
condensation.

In our setup of asymptotically AdS spaces we can apply the
holographic computation of entanglement entropy \RyTa. The entropy
is given by the formula \RyTa \eqn\entropyhol{S_A=\f{{\rm
Area}(\gamma_A)}{4G^{(5)}_N},} where ${\rm Area}(\gamma_A)$ is the
area of the minimal surface $\gamma_A$ whose boundary coincides with
$\de A$. Refer to \Fu\ for its proof from the basic principle of the
AdS/CFT correspondence. Its interpretation from the viewpoint of the
entropy bound (Bousso bound) is given in \HirTa.

First we assume the subsystem $A$ is defined by $x_1>0$ and extends
in the $x_2$ and $\chi$ direction. Then the entropy can be found
from  \entropyhol\ (we put the UV cutoff\foot{Relation to the
lattice spacing introduced in \RyTa\ is given by $r_\infty=R^2/a$.}
$r<r_\infty$)
\eqn\entropyone{S_A=\f{V_1L}{4G^{(5)}_N}\int^{r_\infty}_{r_0}dr
\f{r}{R}= \f{\pi R V_1 r_\infty^2}{8G^{(5)}_N r_0}-\f{\pi RV_1
r_0}{8G^{(5)}_N}.} The first term is divergent and represents the
area law \BKLS \Sr. It is the same as the one in the original
$AdS_5$ background. The second term in \entropyone\
\eqn\secoundt{\Delta S_A=-\f{\pi RV_1 r_0}{8G^{(5)}_N}=-\f{\pi
N^2V_1}{4L},} does not depend on the cut off and thus is physically
important. This is equal to the difference between the entropy in
the AdS bubble and the one in the pure AdS. Since it is negative, we
find that the entropy of the AdS bubble is decreased compared with
the $AdS_5$ solution as we expected. Furthermore, we would like to
conjecture that the AdS bubble has the lowest value among other
asymptotically AdS solution with the same symmetry. This is because
from the energy analysis it is considered to be the lowest energy
configuration \HoMy\ and thus is the most stable solution.

For example, we can consider the time-dependent bubble
 solution\foot{We are very grateful to Gary Horowitz for pointing
 out an important error in this paragraph of the first version of
 our paper.}  found in \HoSi , which has a larger energy. The metric of the constant time
 slice of this solution at $t=0$ is given by
\eqn\otherbubble{ds^2_{t=0}=\left(\f{r^2}{R^2}
-\f{r_0^4}{R^2r^2}\right)d\chi^2+\f{dr^2}
{\left(\f{r^2}{R^2}-\f{r_0^4}{R^2r^2}\right)
\left(1+\f{b}{3r^4-r_0^4}\right)}+\f{r^2}{R^2}(dx_1^2+dx_2^2).} The
specific point $b=0$ is the same as the AdS bubble with the lowest
energy. We have numerically checked that the entropy $S_A$ at $t=0$
computed in the same way as in \entropyone\ for $b>0$
 always takes a larger value than that of the AdS bubble $b=0$.

It is also useful to examine the entropy in the shell configuration
\shell\ since to make sure the existence of tachyon it is better to
start with the shell background of D3-branes. In this case the
entropy can be found from the integration over the codimension three
surface $\gamma_A$, which is similar to the previous one, times
$S^5$ as follows \RyTa
\eqn\entropyshell{\eqalign{S_A&=\f{1}{4G^{(10)}_N}\int_{\gamma_A\times
S^5} dx^8\s{g} \cr &=\f{V_1 L}{4G^{(10)}_N}\int^{r_{\infty}}_0 dr
r^5 h(r)^2 \cr
&=\f{V_1L}{4G_N^{(5)}R}\left[\f{r_{\infty}^2}{2}-\f{r_{0}^2}{3}\right],}}
where $G^{(10)}_N$ is the 10D Newton constant. This is clearly
larger than the entropy in the AdS bubble \entropyone\ and thus the
difference of the entropies again becomes negative
\eqn\secoundtshell{\Delta S_{A(shell)}=-\f{\pi RV_1
r_0}{24G^{(5)}_N}=-\f{\pi N^2V_1}{12L}<0 .} This supports the
conjecture that the AdS shell decays into the AdS bubble.

Now it is also possible to compute $S_A$ when the subsystem $A$ is a
straight belt with a finite width $l$. Suppose $A$ is defined by
$-l/2\leq x_1\leq l/2$, $0\leq x_2\leq V_1(\to \infty)$ and $0\leq
\chi\leq L$. In the dual AdS gravity, we need to consider a minimal
surface $\gamma_A$ whose boundary ( i.e. $r\to \infty$) coincides
with the boundary $\de A$ of $A$.

The area can be written as \eqn\areaofmin{{\rm
Area}=LV_1\int^{l/2}_{-l/2}dx_1
\f{r}{R}\s{\left(\f{dr}{dx_1}\right)^2+\f{r^4f(r)}{R^4}}.} The
energy conservation leads to
\eqn\energyconds{\f{dr}{dx_1}=\f{r^2}{R^2}\sqrt{f(r)\left(
\f{r^6f(r)}{r_*^6f(r_*)}-1\right)},} where $r_*$ is the minimal
value of $r$. The relation between $r_*$ and $l$ is fixed by
\eqn\relationr{\f{l}{2}=\int^\infty_{r_*}dr\f{R^2}{r^2\sqrt{f(r)\left(
\f{r^6f(r)}{r_*^6f(r_*)}-1\right)}}.} Finally the entropy can be
found as \eqn\entropfy{S_A=\f{LV_1}{2RG^{(5)}_N}\int^\infty_{r_*}
\f{r^4\s{f(r)}}{\s{r^6f(r)-r_*^6f(r_*)}}.}

An important point is that when we change the values of $r_*$
arbitrary, only the following specific values of $l$ is allowed by
the relation \relationr\ \eqn\valuechn{l\leq l_0\simeq 0.69\cdot
\f{R^2}{r_0}\simeq 0.22\cdot L.} Thus when $l$ is large enough,
there is no minimal surface that connects the two boundaries of $\de
A$. Under this situation $l>l_0$, the minimal surface is given by
the disconnected sum of the ones considered in \entropyone.

The explicit form of the entropy as a function of $l$ is presented
in Fig.1. We subtracted the entropy $S^{trivial}_A$ for the trivial
disconnected solution (=twice of \entropyone). The lower solution in
Fig.1 is physical compared with the upper one because it has the
lower entropy and gives a dominant contribution to the path-integral
in the gravity. When $S_A-S^{trivial}_A$ becomes positive
($\ti{l}_1\sim 0.31$), the physical solution is replaced by the
trivial disconnected one. Thus there is a phase transition\foot{A
quite similar discontinuity has been found in the holographic
computation of entanglement entropy defined by the annular boundary
in $N=4$ super Yang-Mills \HirTa.} at a specific value $l_1(<l_0)$.
This is not surprising since there will be no correlation between a
distance larger than $\sim L$ due to the mass gap $\sim 1/L$.

\fig{The entanglement entropy as a function of the width $l$ of the
subsystem $A$. We subtracted the entropy $S^{trivial}_A$ for the
trivial disconnected solution (=twice of \entropyone). We plotted
the function $s=s(\ti{l})$ defined by $
S_A-S^{trivial}_A=\f{LV_1r_0^2}{2RG^{(5)}_N}s(l)$ and $\ti{l}=
\f{\pi}{2L}l$. The lower solution is physical compared with the
upper one because it has the lower entropy. }{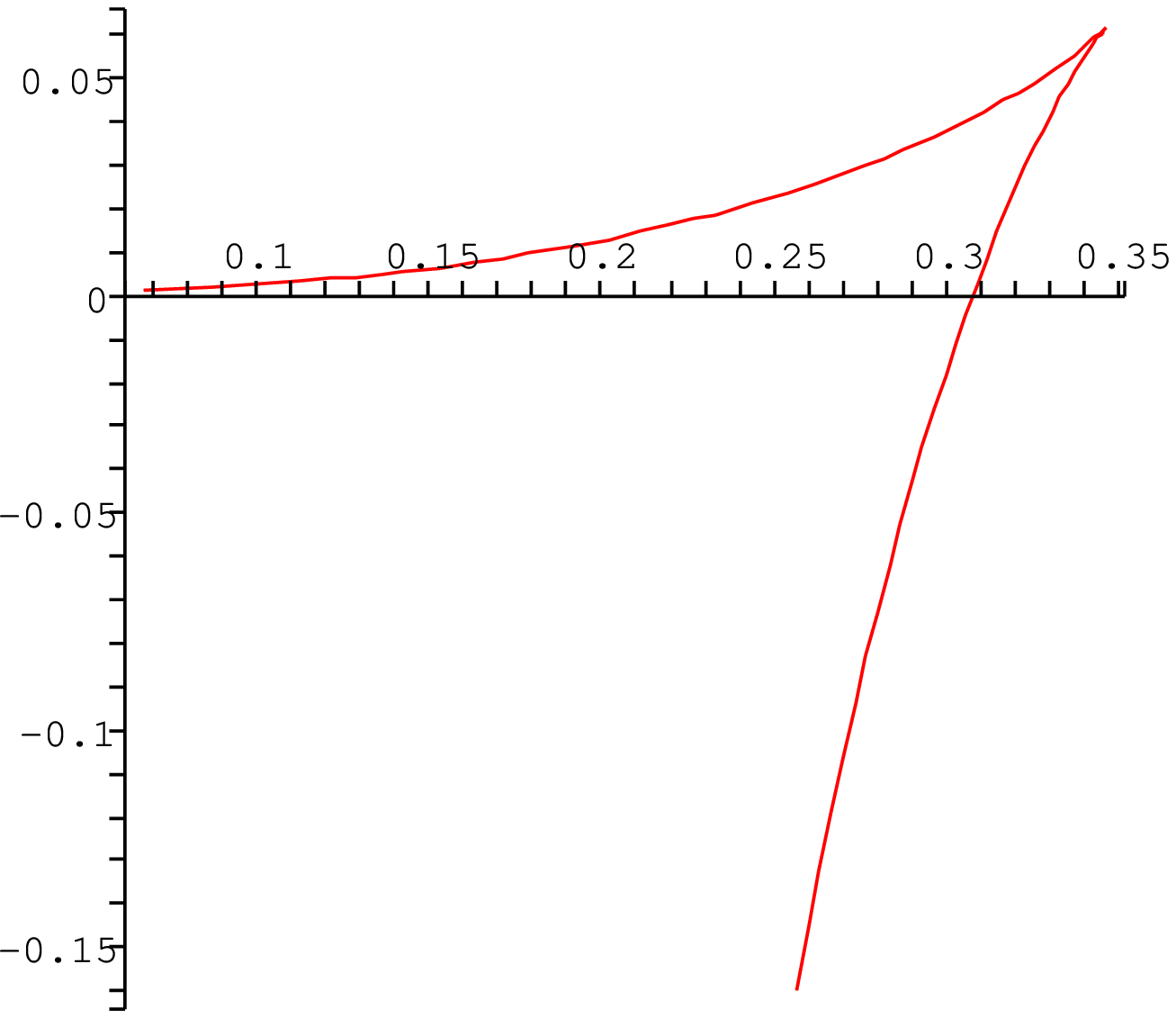}{3truein}

We would also like to notice that the physical solution in Fig.1
(i.e. lower one) is concave as a function of $l$ or equally
$\f{d^2S_A}{dl^2}\leq 0$. This follows from the general property of
the von-Neumann entropy, which is known as the strong subadditivity
\HirTa.

In general 4D conformal field theories, the entanglement entropy
defined in the same way takes the following form \CaHuM \RyTa
\eqn\entrocft{S_A=\gamma\cdot \f{LV_1}{a^2}-C\cdot \f{LV_1}{2l^2},}
where $\gamma$ and $C$ are numerical constants which are
proportional to the number of fields. The first term in \entrocft\
represents the area law divergent term \BKLS \Sr \RyTa. The second
finite term is more interesting because it does not depend on the UV
cutoff $a\to 0$. Motivated by this, we would like to call the
following quantity an entropic c-function
\eqn\entropiccf{C(l)=\f{l^3}{LV_1}\cdot\f{dS_A(l)}{dl}.} This is a
natural generalization of the entropic c-function defined in two
dimension \CaHu \CaHuM. Its explicit form is plotted in Fig.2. Since
$l$ corresponds to the length scale which we are looking at, $C(l)$
measures the degree of freedom at the energy scale $\sim l^{-1}$ in
the given field theory. The Fig.2 shows that its value monotonically
decreases as we decrease the energy scale from the UV region. It
jumps to zero in the middle point and it continues to be vanishing
in the IR region. These are consistent with the expected property of
c-function that it decreases under the RG-flow. The sharp dump of
$C(l)$ is because the IR region $r<r_0$ is completely cutoff in the
AdS bubble solution and represents the clear mass gap in the dual
CFT.

\fig{The form of the entropic c-function. We plotted
$\ti{l}^3\f{ds(\ti{l})}{d\ti{l}}$ as a function of $\ti{l}$. It
jumps to zero at $\ti{l}=\ti{l}_1\sim 0.31$ and for larger values of
$\ti{l}$ it is given by zero.}{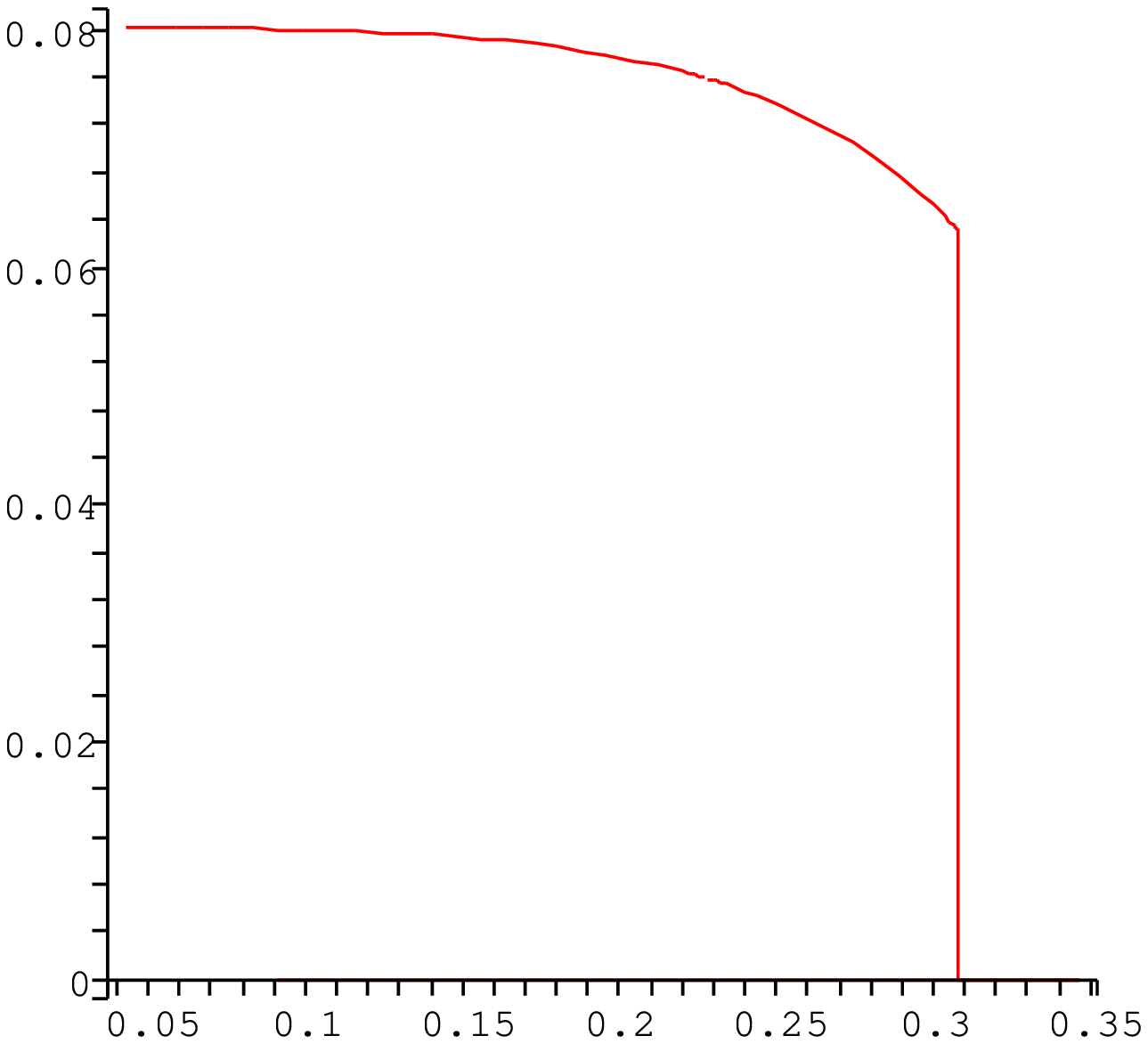}{3truein}

\subsec{Entanglement Entropy: Free Yang-Mills Analysis}

Next we would like to compare the above results from the gravity
side with the direct Yang-Mills free field computations.

Consider a free massless scalar field in $R^{1,d-1}\times S^1$. The
radius of the circle $S^1$ is $L$. We divide the space manifold
$R^{d-1}\times S^1$ (the coordinates are denoted by
$(x^1,x^2,\ddd,x^{d-1};x^d)$) into the submanifolds $A$ and $B$ such
that the boundary $\de A=\de B$ is given by $R^{d-2}\times S^1$
defined by $x_1=0$.

The entanglement entropy can be evaluated as follows (see e.g.\SuUg
\Ka \Cardy \RyTa \Fu). First we compute the partition function
$Z_n=e^{-\beta F}$ on the Euclidean manifold $M_n=\Sigma_n\times
R^{d-2}\times S^1$. The 2D manifold $\Sigma_n$ is the n-sheeted
Riemann surface defined by the metric $ds^2=d\rho^2+\rho^2
d\theta^2$ in the polar coordinate and the conical periodicity
$0\leq \theta \leq \beta=2\pi n$. Then the entropy is obtained by
\eqn\entropyde{S_A=\left[\beta\f{\de}{\de \beta}-1\right](\beta
F)|_{\beta=2\pi},} as if $\beta$ were a real temperature.

One conventional way to compute the free energy $F$ is to employ the
heat kernel method \Ka. Instead, here, we will calculate $F$ using
an orbifold theoretic analysis. In the computation of the entropy,
we can equally consider the positive deficit angle instead of the
negative one. One such example is the orbifold $C/Z_N$. In the setup
of QFT on $R^{1,d}$, the partition function of a free massive scalar
on $C/Z_N\times R^{d-1}$ can be found as
\eqn\partitionorb{\eqalign{\log Z_{C/Z_N}&=V_{d-1}\int^\infty_0
\f{ds}{2s}\cdot \left[\int
\left(\f{dk_\perp}{2\pi}\right)^{d-1}e^{-sk_\perp^2}\right]
\cdot\f{1}{N}\sum_{k=0}^{N-1} \Tr \left[g^{k}\cdot e^{-m^2
s}\right]\cr &=V_{d-1}\int^\infty_0 \f{ds}{2s}\f{1}{(4\pi
s)^{\f{d-1}{2}}}\cdot \f{1}{N}\sum_{k=0}^{N-1} \Tr \left[g^{k}\cdot
e^{-m^2 s}\right].}} This formula is easily obtained by remembering
the expression of the open string cylinder amplitude (see e.g. the
Polchinski's text book \Po) in the Schwinger representation. $\Tr$
in
\partitionorb\ denotes the trace over the zeromodes (the coordinate
$(z,\bar{z})$ and their momenta) of the fields on $R^2(=C)$.
$k_\perp$ are the momenta in the transverse directions $R^{d-1}$.

The summation over $k$ can be done exactly as follows
\eqn\sumksin{\eqalign{\f{1}{N}\sum_{k=0}^{N-1}\Tr
g^{k}&=\f{1}{N}\sum_{k=0}^{N-1}\int dzd\bar{z}\cdot \delta(z-e^{2\pi
ik/N}z) \delta(\bar{z}-e^{-2\pi ik/N}\bar{z})\cr
&=\f{1}{N}\int_{R^2} dz^2\int \f{dk^2}{(2\pi)^2}+\sum_{k=1}^{N-1}
\f{1}{4N\sin^2(\pi k/N)}\cr &={\rm Vol}(C/Z_N)\cdot \int
\f{dk^2}{(2\pi)^2}+\f{1}{12}(N-1/N).}} Thus we can reproduce the
following known expression of the entropy (notice $n=1/N$)
\eqn\entropyexp{\eqalign{S_A&=-\f{\de}{\de(1/N)}\left[\log
Z_{C/Z_N}-\f{\log Z_{C}}{N}\right]_{N=1} \cr &=\f{\de}{\de
N}\left[\f{1}{12}(N-1/N)\cdot V_{d-1}\int \f{ds}{2s}\f{1}{(4\pi
s)^{\f{d-1}{2}}} e^{-m^2s}\right]_{N=1}\cr &=\f{\pi}{3}V_{d-1} \int
\f{ds}{(4\pi s)^{\f{d+1}{2}}} e^{-m^2s}.}}

In our setup, we would like to assume one of the transverse
directions is compactified at the radius $\f{L}{2\pi}$. Then we
obtain \eqn\final{S_A=\f{\pi}{3}V_{d-2}\int^\infty_{a^2}\f{ds}{(4\pi
s)^{d/2}}\cdot
\f{L}{2\pi}\cdot\s{\f{\pi}{s}}\cdot\sum_{q=-\infty}^\infty
e^{-\f{L^2q^2}{4s}},} where we introduced the UV cutoff (or the
lattice spacing) $a$. We divide \final\ into the divergent $q=0$
term and the finite $q\neq 0$ term
\eqn\divided{S_A=S^{area~law}_A+S^{finite}_A,}
\eqn\arealawt{S^{area~law}_A=\f{\pi}{3}\cdot L
V_{d-2}\cdot\int^{\infty}_{a^2} \f{ds}{(4\pi s)^{(d+1)/2}},}
\eqn\finite{S^{finite}_A=\f{V_{d-2}}{3}\cdot
2^{-1}\pi^{\f{1}{2}-\f{1}{2}d}\cdot
\Gamma\left(\f{d}{2}-\f{1}{2}\right)\cdot \zeta(d-1)\cdot
\f{1}{L^{d-2}}.}

In particular, the 4D massless scalar $d=3$, we find
\eqn\finite{S^{finite}_A=\f{\pi V_1}{36L}.}

Now we turn to free fermions. By direct computation we can show that
the expression of $S_A$ is the same form and its coefficient is
proportional to the central charge $c$ when we reduce the system to
two dimension \Ka. One way to understand this is to note that the
entropy of the $d+1$ dimensional free field theory is obtained from
the two dimensional entropy with the correlation length $\xi=1/m$
\Cardy\ \eqn\twod{S_A=\f{\pi c}{3}\int^\infty_{a^2}\f{ds}{4\pi
s}e^{-sm^2} =\f{c}{6}\log(ma),} by summing over the KK modes as
follows \eqn\summas{\eqalign{S_A&=\f{\pi c}{3}V_{d-2}\int
\left(\f{dk_{\bot}}{2\pi}\right)^{d-1} \int^\infty_{a^2}\f{ds}{4\pi
s}e^{-s(m^2+k_{\bot}^2)} \cr &=\f{\pi
c}{3}V_{d-2}\int^\infty_{a^2}\f{ds}{(4\pi s)^{(d+1)/2}}e^{-sm^2} .}}

Thus for each real component of a Majorana fermion with the periodic
boundary condition we find \eqn\entropp{
S_A^{(P)}=\f{\pi}{6}V_{d-2}\int^\infty_{a^2}\f{ds}{(4\pi
s)^{d/2}}\cdot
\f{L}{2\pi}\cdot\s{\f{\pi}{s}}\cdot\sum_{q=-\infty}^\infty
e^{-\f{L^2q^2}{4s}}.} On the other hand, the anti-periodic fermion
leads to \eqn\entroppa{
S_A^{(A)}=\f{\pi}{6}V_{d-2}\int^\infty_{a^2}\f{ds}{(4\pi
s)^{d/2}}\cdot
\f{L}{2\pi}\cdot\s{\f{\pi}{s}}\cdot\sum_{q=-\infty}^\infty
(-1)^q\cdot e^{-\f{L^2q^2}{4s}}.} When $d=3$, the finite part of the
entropy for each of them is given by \eqn\entropyr{\eqalign{&
S^{finite(P)}_A=\f{\pi V_1}{72L}, \cr & S^{finite(A)}_A=-\f{\pi
V_1}{144L}.}}

In summary, the finite part of the entanglement entropy in N=4
$SU(N)$ super Yang-Mills theory can be found in both periodic and
anti-periodic fermion cases as follow \eqn\nfourym{
S^{finite(P)}_A=\f{N^2\pi V_1}{3L},\ \ \ \ \ \
S^{finite(A)}_A=\f{N^2\pi V_1}{6L}.} Their difference \eqn\differs{
\Delta S_A =S^{(A)}_A-S^{(P)}_A=-\f{\pi N^2V_1}{6L},} should be
compared with the AdS result \secoundt. They differs only by the
factor $\f{2}{3}$. Again, we can think this a successful agreement
since the gravity calculation is dual to the strongly coupled
Yang-Mills, while our gauge theoretic result is found for the free
Yang-Mills. The entanglement entropy is not protected by any
supersymmetries because the conical geometry appears in the
definition \entropyde\ breaks all of supersymmetries.

Finally it may be interesting to examine the entanglement entropy in
the free Yang-Mills when the subsystem $A$ is defined by the
straight belt with a finite width as in \entrocft. This has been
done in \RyTa\ for the N=4 super Yang-Mills and the entropy of the
form \entrocft\ was obtained. The strategy is to first regard the
system to infinitely many 2D free field theories and then to
integrate the known numerical results  of the 2D entropic c-function
\CaHuM. We can repeat the same computation in our compactified
Yang-Mills theory. However, it does not seem to be possible to
reproduce the phase transition as a function $l$ found in the
previous gravity analysis. This will be essentially because the
phase transition occurs due to the strongly coupled phenomena (i.e.
confinement), while we are treating the free Yang-Mills.

\newsec{Twisted AdS Bubbles and Closed String Tahcyons}

We would like to study the second example i.e. the double Wick
rotation of the rotating non-extremal D3-branes \Gu \Ru \Cv. We will
claim that this solution (we call it the twisted AdS bubble\foot{
Refer to \BiDeMu\ for the time-dependent bubble solution obtained by
another double Wick rotation of R-charged solutions in 5D gauged
supergravity.}) is the end point of the decay of the D3-brane
background with a twisted boundary condition (i.e. the twisted
circle or Melvin background \RuTe \DaGuHeMi). By taking the near
horizon limit, this is equivalent to the statement that the AdS with
the twisted identification (we call it the AdS twisted circle) will
decay into the twisted AdS bubble via closed string tachyon
condensation.

\subsec{Twisted AdS Bubble Solution}

After the double Wick rotation $t\to i\chi$, $\chi\to it$ and $l\to
-il$ of the rotating black 3-brane solution \Gu \Ru \Cv, the
solution looks like \eqn\rotate{\eqalign{ ds^2
&=\f{1}{\s{f}}(-dt^2+h\
d\chi^2+dx_1^2+dx_2^2)+\s{f}\Bigl[\f{dr^2}{\ti{h}}
-\f{2lr_0^4\cosh\ap}{r^4\Delta f}\sin^2\theta d\chi d\phi\cr &\
+r^2(\Delta d\theta^2+\ti{\Delta}\sin^2\theta d\phi^2+\cos^2\theta
d\Omega_3^2)\Bigr],}} where $f$, $h$, $\ti{h}$, $\Delta$ and
$\ti{\Delta}$ are defined as follows \eqn\defhf{\eqalign{ &
f=1+\f{r_0^4\sinh^2\ap}{r^4\Delta},\ \
\Delta=1-\f{l^2\cos^2\theta}{r^2},\ \
\ti{\Delta}=1-\f{l^2}{r^2}-\f{r_0^4l^2\sin^2\theta}{r^6\Delta f},\cr
& h=1-\f{r_0^4}{r^4\Delta},\ \
\ti{h}=\f{1-\f{l^2}{r^2}-\f{r_0^4}{r^4}}{\Delta}.}} The parameter
$l$ before the double Wick rotation is proportional to the angular
momentum of the black brane solution.

The allowed lowest value $r_H$ of $r$ is given by the solution to
$\ti{h}(r)=0$ \eqn\rhiv{r_H^2= \f{l^2}{2}+\s{r_0^4+\f{l^4}{4}}\ \ \
(>l^2).} It is easy to see that $f$, $h$, $\ti{h}$, $\Delta$ and
$\ti{\Delta}$ are all positive when $r>r_H$. The total 3-brane
RR-charge is proportional to $r_0^4\cosh\ap\sinh\ap(\propto N)$ and
it is taken to be a finite constant.

 If we set the
angular momentum to zero $l=0$, then it is reduced to the previous
example of the AdS bubble. In the near horizon limit, we can
approximate it as $f\simeq \f{r_0^4\sinh^2\ap}{r^4\Delta}$ and thus
the AdS radius $R$ is given by $R^2=r_0^2\sinh\ap$. Thus we call
this solution the twisted AdS bubble.

As in the previous example we need to be careful about the
regularity of the solution. The non-trivial constraint comes from
the behavior near the point $\theta=0$ and $r=r_H$. Around that
point, the relevant part of the metric looks\foot{Notice
$\Delta=\ti{\Delta}=\f{r_0^4}{r_H^4}$ and $f=\cosh^2\ap$.}
\eqn\metricex{\eqalign{ds^2\simeq &\f{\beta
r_H^4}{r_0^4\cosh\ap}(r-r_H) d\chi^2+\f{r_0^4\cosh\ap }{\beta
r_H^4}\f{dr^2}{r-r_H}\cr &+\f{r_0^4\cosh\ap }{r_H^2}\left[d\theta^2
+\theta^2(d\phi-\f{l r_H^2}{r_0^4\cosh\ap}d\chi)^2\right],}} where
$\beta=\f{4}{r_H}-\f{2l^2}{r_H^3}$.

The regularity requires the following two identifications
\eqn\periodrota{\eqalign{&(\chi,\phi)\sim (\chi,\phi+2\pi),\cr &
(\chi,\phi)\sim (\chi+L,\phi+2\pi\zeta),}}where \eqn\leng{
L=\f{2\pi r_0^4\cosh\ap}{2r_H^3-l^2r_H},\ \ \  \zeta=
\f{lr_H^2}{2r_H^3-l^2r_H}.}

\subsec{Twisted Circle Background}

The second condition in \periodrota\ looks non-trivial. Though in
the asymptotic region $r\to \infty$, the form of the metric
approaches the flat metric \eqn\flatmet{ds^2=
-dt^2+d\chi^2+dx_1^2+dx_2^2+dr^2+r^2(d\theta^2+\sin^2\theta
d\phi^2+\cos^2\theta d\Omega_3^2),} the second periodicity requires
that $(\chi,\phi)$ is identified $(\chi,\phi)\sim
(\chi+L,\phi+2\pi\zeta)$. This means that the asymptotic geometry is
the twisted circle (or Melvin background). The string theory on such
a background was first studied in \RuTe.

The parameter $\zeta$ which measures the strength of the twist takes
the values within $0\leq \zeta <1$. This is clear if we rewrite it
as follows \eqn\zett{\zeta=\s{\f{x^4+x^2\s{4+x^4}}{2(x^4+4)}},\ \ \
\left(x\equiv \f{l}{r_0}\right).} The upper bound $\zeta <1$ is very
natural since the point $\zeta=1$ corresponds to the supersymmetric
compactification in the asymptotic region and thus there should be
no bubble solution. Indeed we can see that the limit $\zeta\to 1$ is
equivalent to the extremal D3-branes $l,r_0\to 0$, keeping $L$ and
$N$ finite.

Sometimes it is useful to define the new angular coordinate
$\ti{\phi}=\phi-q\chi$, where $qL=2\pi\zeta$, and rewrite the metric
\flatmet\ as follows \eqn\rewrite{ds^2=
-dt^2+d\chi^2+dx_1^2+dx_2^2+dy^2+y^2(d\ti{\phi}+q
d\chi)^2+\sum_{i=1}^4 dz_i^2,} where we used the transverse
coordinates defined by $y\equiv r\sin\theta$ and $z_i\equiv
r\cos\theta({\Omega_3})_i$. Notice that in this new coordinate
system, the periodicity of $\ti{\phi}$ and $\chi$ are are given by
the ordinary (untwisted) ones $\ti{\phi}\sim \ti{\phi}+2\pi$ and
$\chi\sim \chi+L$.

The string theory on the twisted circle does not change continuously
with respect to the parameter $\zeta$. Especially, when $\zeta$ is
irrational, it is known that the string theory behaves rather
unusually \TaUe \KuMaMo. Thus below we will mainly assume that
$\zeta$ takes rational values \eqn\rational{\zeta=\f{k}{M},} where
$k$ and $M$ are coprime positive integers. In this case the string
theory on \rewrite\ is equivalent to the one on the $Z_M$ orbifold
$(R^2\times S^1)/Z_M\times R^{1,6}$. To be more precise, the
background before the $Z_M$ projection is considered to be the
ordinary supersymmetric type II string when $k+M$ is even. On the
other hand, when $k+M$ is odd, it is the type II string with an
antiperiodic boundary condition for fermions in the circle
direction\foot{In other words, it is the $Z_2$ orbifold of type II
string by the action $\sigma_{1/2}\cdot (-1)^{F_S},$ where
$\sigma_{1/2}$ is the half shift along the circle and $F_S$ is the
spacetime fermion number.}. If we take the small radius limit $L\to
0$, the background becomes equivalent to the type II string (or type
0 string) on $R\times (C/Z_M)$ when $k+M$ is even (or is odd) \TaUe.

\subsec{Closed String Tachyon Condensation}

It is known that in the twisted circle background (or Melvin
background) of type II string, a closed string tachyon appears when
the radius of circle is enough small \RuTe\ ($~$string scale). This
tachyon is localized near the origin of $R^2$ \RuTe \CoGu \SuMe
\TaUe \RTF \DaGuHeMi. Suppose $N$ D3-branes are located at $r=0$ of
the twisted circle \flatmet. Then its near horizon metric is given
by \eqn\adsdthtw{ds^2=R^2
\f{dr^2}{r^2}+\f{r^2}{R^2}(-dt^2+d\chi^2+dx_1^2+dx_2^2)+
R^2(d\theta^2+\sin^2\theta d\phi^2+\cos^2\theta d\Omega_3^2).}
 with the
identification $(\chi,\phi)\sim (\chi+L,\phi+2\pi\zeta)$. Then we
find that the radius of the twisted circle $\chi$ becomes small in
the IR region $r<<1$. Thus we expect the closed string tachyon
condensation in that region. We would like to argue that the end
point is given by the twisted bubble \rotate\ in the similar sense
of the previous example of the AdS bubble. To make this argument
clearer, we can assume the shell distribution of the D3-branes so
that the flat spacetime is realized inside the shell as before. We
will presents non-trivial evidences by comparing the energy density
and the entanglement entropy between the gauge and gravity side in
the following subsections.

To see if this speculation makes sense, it is useful to see how the
twisted boundary condition \periodrota\ at the UV boundary
$r=\infty$ evolves toward the IR region $r\to r_0$. If we rewrite
the metric \rotate\ near $\theta=0$ in the form $A(d\chi)^2+
B\theta^2(d\phi-q(r)d\chi)^2+..$, the twist parameter $q(r)$ is
given by \eqn\qpa{q(r)=\f{lr_0^4\cosh\ap}{\Delta^2 f r^6}.} This
becomes monotonically large toward the IR region and it becomes zero
in the UV limit $r=\infty$. Since we put the twisted boundary
condition at $r=\infty$ and the non-zero value of $q(r)$ cancels the
effect of the twist, the strength of the twist becomes weaker as we
go into the IR region and it vanishes at $r=r_H$ smoothly. This is
qualitatively consistent with the fact that the closed string
tachyon condenses only in the IR region, where the radius of the
circle becomes stringy size and that the UV geometry should not
change.

At the same time, another important property of the closed string
tachyon in the AdS twisted circle is that it is localized near the
$S^3$ (`north pole') defined by $\theta=0$ within the whole $S^5$.
This is the crucial difference between this example and the previous
one in section 2. Indeed, the geometry depends on the position of
$S^5$ as is clear from the metric \rotate. We can find that the
radius of the twisted circle (defined by the shift $\Delta\chi=L,\
\Delta\phi=2\pi\zeta$) depends on $\theta$ and it is non-zero except
the north pole $S^3$. Thus the twisted circle shrinks to zero size
in the north pole. This is consistent with the fact that the tachyon
is localized at the north pole and the observation in \ALMSS\ that
the winding tachyon will pinch off the wound circle.

\subsec{Asymptotically Flat Solution with a Conical Singularity}

In the previous subsection the background asymptotically approaches
the twisted circle and thus is not asymptotically flat. Instead, we
can consider the same solution \rotate\ with requiring the
asymptotic flatness. Inevitably, we will encounter conical
singularities in the IR region at $\theta=0,r=r_H$. Notice that we
consider this asymptotically flat solution only in this subsection
among all parts of the present paper.

Let us first examine what types of the singularities appear in the
IR region. The asymptotic flatness requires the coordinates $\phi$
and $\chi$ are compactified in a usual way i.e. \eqn\compactr{
(\chi,\phi)\sim (\chi+L',\phi),\ \ \ \ (\chi,\phi)\sim
(\chi,\phi+2\pi),} where the periodicity $L'$ is not necessarily
equal to $L$ in \leng. Now we assume the combination $\f{\zeta
L'}{L}$ is a rational number and we express it as $\f{\zeta
L'}{L}=\f{k}{M}$, where $M$ and $k$ are coprime integers. Define the
following two angles \eqn\defanb{
\ti{\chi}\equiv\f{2\pi\zeta}{kL}\chi,\ \ \
\ti{\phi}=\phi-\f{2\pi\zeta}{L}\chi.} They satisfy the following
periodicity \eqn\periodchiphi{(\ti{\chi},\ti{\phi})\sim
(\ti{\chi}+\f{2\pi}{M},\ti{\phi}-\f{2\pi k}{M}),\ \ \ \
(\ti{\chi},\ti{\phi})\sim (\ti{\chi},\ti{\phi}+2\pi).} Thus if we
define the following coordinate of $R^4=C^2$ in the neighborhood of
$\theta=0,r=r_*$ (irrelevant constant factors are denoted by $a$ and
$b$)\eqn\defcordc{Z_1=a \s{r-r_*} e^{i\ti{\chi}},\ \ \ Z_2=b\theta
e^{i\ti{\phi}},} the singular geometry is described by the orbifold
$C^2/Z_M$ described by the $Z_M$ action \eqn\actionor{ (Z_1,Z_2)\sim
(Z_1 e^{\f{2\pi i}{M}},Z_2 e^{-\f{2\pi ik}{M}}).}

Next we would like to estimate the energy of these configurations.
It is known that the ADM energy of the rotating black hole is given
by the same formula as the non-rotating one i.e. $l=0$ \Ru \Gu. This
is because its asymptotic geometry $r\to \infty$, where we read off
the ADM mass, does not depend on the parameter $l$ \HaObT \HaOb. The
same is true for our case since the double Wick rotation  does not
essentially touch\foot{Even though the off diagonal term $\propto
dtd\phi$ depends on $l$, its coefficient becomes too small to
contribute to the ADM mass when $r$ is large.} the terms which
depend on $l$.

Thus we obtain the energy density of the twisted bubble
 \eqn\energyy{T_{00} =\f{\pi^2r_0^4}{16
G^{(10)}_N}(1+4\sinh^2\ap).} We would like to subtract the energy
stress tensor of extremal D3-branes from \energyy. The extremal
limit is given by $r_0\to 0$ (or $\ap\to \infty$). To do this
subtraction we should keep the total RR-flux same in both sides. The
RR-flux is proportional to $r_0^4\sinh\ap\cosh\ap$ as we mentioned.
The energy density of extremal D3-branes with the same amount of
RR-flux is \eqn\extremalt{T^{(0)}_{00}=\f{\pi^2
r_0^4}{4G^{(10)}_N}\sinh\ap\cosh\ap.} After the subtraction the
energy density becomes
\eqn\energydensu{T_{00}-T^{(0)}_{00}=-\f{\pi^2r_0^4}{16G^{(10)}_N}.}
Even though this expression looks equivalent to the AdS bubble
($l=0$), its physical value depends on $l$ non-trivially via the
relation \leng\ (we always fix the value of $L$).

\subsec{Casimir Energy}

Now we come back to the geometry \rotate\ with the twisted
identification \periodrota. It is dual to the $SU(N)$ Yang-Mills
theory with the twisted boundary condition. This originates from the
D3-branes wrapped on the circle $S^1$ in the orbifold $(R^2\times
S^1)/Z_M$ \TaUeD \HeMiTa (see also \MiYi). Let us compute the
Casimir energy in this gauge theory.

The transverse (complex) scalar in the $R^2$ direction is denoted by
$\Phi$, and the other scalars are denoted by $\phi$. Their twisted
boundary conditions are written as
\eqn\twist{\Phi_{ab}(z+L)=e^{\f{2\pi i}{M}(a-b+k)}\Phi_{ab}(z),\ \ \
\ \phi_{ab}(z+L)=e^{\f{2\pi i}{M}(a-b)}\phi_{ab}(z),} where $a$ and
$b$ represents the Wilson line in the circle direction and take
values\foot{ As shown in \TaUeD, the values $i\leq a,b<i+1$
corresponds to the $i$-th fractional branes in the orbifold limit
$L\to 0$.} $0\leq a,b<M$. For the fermions we similarly find
\eqn\twist{\psi_{ab}(z+L)=e^{\f{\pi i(k+M)}{M}}e^{\f{2\pi
i(a-b)}{M}}\psi_{ab}(z).}

The two point functions for these fields can be found easily. For
example, the one for the field $\phi$ becomes
\eqn\correlationsct{\la \phi(x)_{ab}\phi(x')_{ba}\lb
=\f{1}{4\pi^2}\sum_{n\in {\bf Z}}\f{e^{\f{2\pi
i}{M}(a-b)n}}{(x-x')^2+(y-y')^2+(z-z'-nL)^2-(t-t')^2}.} As in
section 2.2, it is straightforward to compute the Casimir energy
from \correlationsct. When we consider $N$ D3-branes with the same
value of the Wilson line, we obtain \eqn\casike{\eqalign{T_{00}&
=\f{N^2}{\pi^2 L^4}\left[-6\sum_{n=1}^\infty
\f{1}{n^4}-2\sum_{n=1}^\infty \f{\cos\bigl(\f{2\pi
nk}{M}\bigr)}{n^4}+8\sum_{n=1}^\infty \f{\cos\bigl(\f{\pi
n(k+M)}{M}\bigr)}{n^4} \right] \cr &= -\f{N^2\pi^2}{L^4}
\left[\f{1}{6}-\f{k^2}{M^2}+\f{4k^3}{3 M^3}-\f{k^4}{2M^4} \right]\cr
&=-\f{N^2\pi^2}{L^4} \left[\f{1}{6}-\zeta^2+\f{4}{3}\zeta^3
-\f{1}{2}\zeta^4 \right],}} where we have employed the identity
\eqn\formulasum{\sum_{n=1}^\infty\f{\cos(nx)}{n^4}=\f{1}{48}\left[
2\pi^2(x-\pi)^2-(x-\pi)^4-\f{7}{15}\pi^4\right].}

 This energy density \casike\ is a
monotonically increasing function of $\zeta$ (see the lower graph in
Fig.3). In particular, it takes the vanishing value $T_{00}=0$ at
the supersymmetric point $\zeta=1$ and the previous value \nfouren\
at $\zeta=0$. Notice also that $T_{00}$ is negative except the
supersymmetric point, which is consistent with our claim that the
closed string tachyon condensation leads to the twisted AdS bubble.

In the above we assumed that D3-branes at the twisted circle
$(R^2\times S^1)/Z_M$ have the same value of the Wilson line $a$. In
the orbifold theoretic language, such branes are called fractional
D3-branes of the same type. We cannot move them away from the origin
of $R^2$ without exciting the system \TaUeD \DuMo.

It is also intriguing to consider a bulk D3-brane, which is
equivalent to a linear combination of $M$ fractional D3-branes of
different types. A bulk D3-brane has a moduli which shifts its
position away from the origin. To compute the Casimir energy of $N$
bulk D3-branes we need to sum over $a$ and $b$ such that
$a,b=0,1,2,\ddd,M-1$. This can be easily done because
$\sum_{a,b}e^{\f{2\pi i}{M}(a-b)n}=M^2\cdot\delta_{n,M{\bf Z}}.$
Thus the total Casimir energy is given by replacing $L$ with $ML$
and multiplying $M^2$. Also we have to be careful about the boundary
condition for fermions. In the end we find the following result:
when $k+M$ is even, the energy is vanishing, while $k+M$ is odd, it
is given by \eqn\totaltwist{T_{00}=-\f{\pi^2 N^2}{6M^2L^4}.}

It is now clear that the system of bulk D3-branes has a larger
energy compared with that of the fractional D3-branes and thus it is
unstable. These results for the bulk D3-branes also tell us that the
dual background is given by the $Z_M$ orbifold of the pure AdS or of
the (untwisted) AdS bubble with the periodicity $\chi\sim\chi+ML$
when $k+M$ is even or odd, respectively. Thus when the $k+M$ is odd,
the bulk tachyon is condensed in the IR region, while $k+M$ is even,
closed string tachyons is not condensed.

In other words, the tachyon condensation process from the AdS
twisted circle to the AdS twisted bubble is dual to the shift of the
Wilson line expectation values\foot{This means that the tachyon
condensation corresponds to the clumped eigenvalues of Wilson loop.
This looks analogous to the behavior of the 2D maximally
supersymmetric Yang-Mills pointed out in the paper \AMMW, which
relates the clumping phenomena to the Gregory-Laflamme
black-hole/black string transition. It is also similar to the free
Yang-Mills analysis \Sud \Ah\ of the deconfinement phase transition
\Wi.} from those for the bulk D3-branes to those for the fractional
D3-branes if $k+M$ is even.

\vskip .3in

To make the above point clearer, let us compare the energy computed
in the free Yang-Mills with the one \energydensu\ found in the
gravity side. Here we have to be careful since we are treating the
energy in a background which is not asymptotically flat or AdS.
Nevertheless, we assume the result \energydensu, which was obtained
by requiring the asymptotic flatness, is also true for our case with
the twisted boundary condition. This is reasonable because the
twisted circle is a freely acting orbifold and will not produce any
extra energy as opposed to the conical orbifold $C/Z_n$ \HaHo
\GuHeMiSc.

The energy density in the gravity side reads in terms of the gauge
theoretic variables \eqn\energyadms{T_{00}=-\f{\pi^2N^2}{8L^4}
\cdot\f{1}{(x^2/2+\s{1+x^4/4})^2(1+x^4/4)^2}\simeq -\f{\pi^2
N^2}{L^4}\left(\f{1}{8}-\f{1}{2}\zeta^2+\ddd\right),} where in the
final expression we wrote down the power expansion with respect to
$\zeta$. This result is plotted as a upper graph in Fig.3. It takes
the values $T_{00}\to -\f{\pi^2N^2}{8L^4}\ (\zeta\to 0)$ and
$T_{00}\to 0\ (\zeta\to 1)$. The qualitative behavior of the
gravitational energy agrees with the Casimir energy in free
Yang-Mills \casike\ rather successfully as is clear from Fig.3. The
ratio of the energy density in both sides is given by
$\f{T^{freeYM}_{00}}{T^{gravity}_{00}}=\f{4}{3}$ for the AdS bubble
$\zeta=0$ as we have already reviewed in section 2.2. A new result
here is that in the extremal limit $\zeta\to 1$ it approaches
\eqn\ratio{\f{T^{freeYM}_{00}}{T^{gravity}_{00}}=\f{9}{8}.} This
value is closer to $1$ than the result at $\zeta=0$, which is very
natural because the free Yang-Mills can be a better approximation to
the strongly coupled Yang-Mills in the almost BPS case than in the
deeply non-BPS case.

\fig{The Casimir energy (in the normalization of
$\f{L^4}{\pi^2N^2}\cdot T_{00}$) as a function of the twist
parameter $0\leq \zeta\leq 1$ is presented in both gravity and free
Yang-Mills side. The one starts with the value $-0.125$ is the
gravity result and the other is the free Yang-Mills
result.}{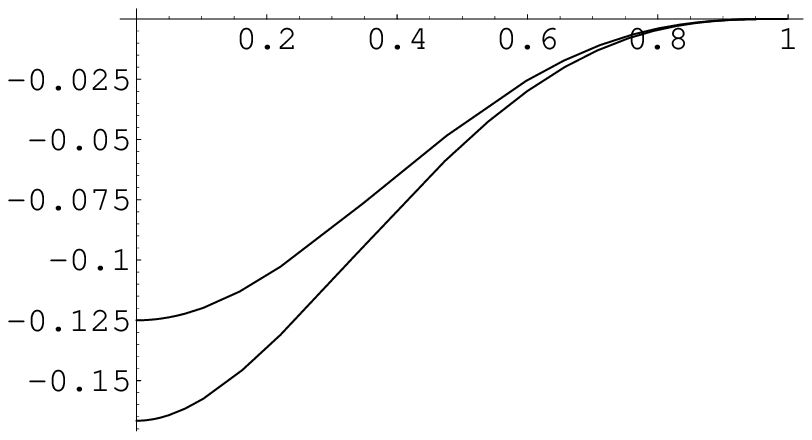}{3truein}

\subsec{Entanglement Entropy}

The entanglement entropy in the free $U(N)$ N=4 Yang-Mills theory
can be obtained as in section 2.4. The result for the field with the
twisted boundary condition $\phi(z+2\pi R)=e^{\f{2\pi
i}{M}a}\phi(z)$ is obtained from the untwisted one by replacing the
sum $\sum_{q\in Z}e^{-\f{\pi^2 R^2 q^2}{s}}$ with $\sum_{q\in
Z}e^{\f{2\pi i}{M}aq}\cdot e^{-\f{\pi^2 R^2 q^2}{s}}$ in \final. We
can perform the summation by using the formula \eqn\formulasumm{
\sum_{n=1}^\infty\f{\cos(nx)}{n^2}=\f{1}{4}(x-\pi)^2-\f{\pi^2}{12}.}

Suppose $N$ fractional D3-branes on the twisted circle. The entropy
can be found as \eqn\entropyfr{ S_A=12N^2\cdot
S^{area~law}_A+\f{N^2\pi
V_1}{6L}\left(1+\f{3k^2}{M^2}-\f{2k}{M}\right),} where
$S^{area~law}_A$ is defined by \arealawt. Thus we obtain
\eqn\entropyfrs{ \Delta S_A^{freeYM}=-\f{N^2\pi
V_1}{6L}\left(1+2\zeta-3\zeta^2\right).} Since this is clearly
negative $\Delta S_A<0$ (for $0\leq \zeta<1$), we again confirm that
the entanglement entropy decreases under the closed string tachyon
condensation.

In the case of $N$ bulk D3-branes, the entropy takes the following
form (we assume $k+M$ is odd) \eqn\entropytw{ S_A=12M^2N^2\cdot
S^{area~law}_A+\f{N^2\pi V_1}{6L},} where $S^{area~law}_A$ is
defined by \arealawt. Thus the finite term does not depend on $M$.
When $k+M$ is even, the second finite term is given by $\f{N^2\pi
V_1}{3L}$. Again these results are consistent with the previous
claim that the dual background of the bulk 3-branes is given by the
orbifold of the pure AdS or of the (untwisted) AdS bubble with the
periodicity $ML$ when $k+M$ is even or odd, respectively, by
considering the entropy density $\f{\Delta S_A}{LV_1}$.

\vskip .2in

Next we compare these with the gravity calculation. We would like to
apply the holographic computation of the entanglement entropy to the
near horizon limit $e^\ap>>1$ of the twisted bubble \rotate. In this
example, the total 10D spacetime is relevant and thus we need to
apply the  holographic formula generalized into ten dimension \RyTa
\eqn\formulaten{S_A=\f{1}{4G^{(10)}_N}\int_{\gamma_A\times
S^5}\s{g},} which was already employed in \entropyshell. We only
consider the simplest case where the subsystem is defined by
dividing the total space into half parts as in \entropyone. After
some algebras we find that the integral in \formulaten\ becomes
drastically simplified as
\eqn\entrotwist{S_A=\f{V_1L}{4G^{(5)}_N}\int^{r_{\infty}}_{r_H} dr
\f{r}{R}
=\f{V_1L}{4G^{(5)}_NR}\left(\f{r_{\infty}^2}{2}-\f{r_H^2}{2}\right).}
In terms of the gauge theoretic language assuming $\zeta=\f{k}{M}$,
the second finite term is equivalently rewritten as ($x=l/r_0$)
\eqn\entyroymtw{\Delta S_A^{gravity}= -\f{\pi N^2V_1}{(4+x^4)L}.}
Again we confirmed $\Delta S_A<0$ and this
 agrees with our conjecture.

The Fig.4. summarizes the results in both free Yang-Mills (upper)
and gravity side (lower). First we notice that in the near extremal
region $1-\zeta\ll 1$, the entropy coincides precisely
\eqn\agreeens{\Delta S_A^{freeYM}\simeq \Delta S_A^{gravity}
\simeq\f{2\pi N^2V_1}{3L}(\zeta-1).} On the other hand, near
$\zeta=0$, their behaviors are slightly different. The free
Yang-Mills entropy takes a minimum value at $\zeta=\f{1}{3}$, while
the gravity (or strongly coupled gauge theory) entropy is a
monotonically increasing function. Since the entanglement
 entropy is not protected by supersymmetries, we may have to
 be satisfied with this result, though it is not clear why the free
 Yang-Mills entropy takes the minimum value. The quantitative
 agreement in the near extremal region \agreeens\ is remarkable
 from this conventional viewpoint and may be regarded as
 a further evidence for AdS/CFT correspondence in slightly
 non-BPS backgrounds.
\fig{The entanglement entropy (in the normalization of $\f{L}{\pi
N^2 V_1}\Delta S_A$) as a function of the twist parameter $0\leq
\zeta\leq 1$ is presented in both gravity and free Yang-Mills side.
The one starts with the value $-0.25$ is the gravity result and the
other is the free Yang-Mills one.}{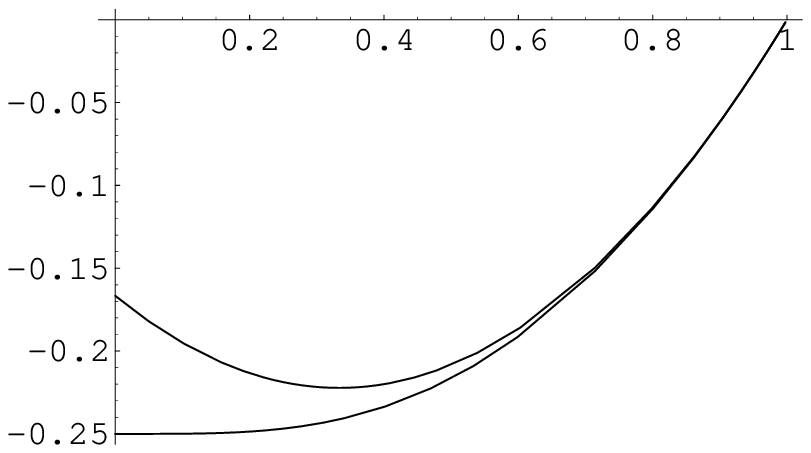}{3truein}

\subsec{More General Solutions}

We can construct more general twisted AdS bubble solutions by the
double
 Wick rotation of rotating D3-brane solutions \Cv\ with three angular momenta
 (or equally three R-charges)
 \eqn\rotatetwo{\eqalign{ ds^2
&=\f{1}{\s{f}}(-dt^2+h\
d\chi^2+dx_1^2+dx_2^2)+\s{f}\Bigl[\f{dr^2}{\ti{h}}+
r^{2}\sum_{i=1}^{3}H_{i}(d\mu_{i}^{2}+\mu_{i}^{2}d\phi_{i}^{2})\cr
&\ -\f{2r_0^4\cosh\ap}{r^4\Delta
f}d\chi(\sum_{i=1}^{3}l_{i}\mu_{i}^{2} d\phi_{i})
-\f{r_0^4}{r^4\Delta f}(\sum_{i=1}^{3}l_{i}\mu_{i}^{2}
d\phi_{i})^{2} \Bigr],}}
 where $f$, $h$, $\ti{h}$, $\Delta$ and
$H_{i} (i=1,2,3)$ are defined as follows \eqn\defhftwo{\eqalign{ &
f=1+\f{r_0^4\sinh^2\ap}{r^4\Delta},\ \ H_{i}=1-\f{l_{i}^2}{r^2},\ \
\Delta=H_{1}H_{2}H_{3}\sum_{i=1}^{3}\f{\mu_{i}^{2}}{H_{i}},\cr &
h=1-\f{r_0^4}{r^4\Delta},\ \
\ti{h}=\f{H_{1}H_{2}H_{3}-\f{r_0^4}{r^4}}{\Delta},}} and
\eqn\coord{(\mu_{1}, \mu_{2}, \mu_{3})=(\sin\theta,
\cos\theta\sin\phi, \cos\theta\cos\phi).} If we set $l_{2}=l_{3}=0$,
this metric is reduced to the previous example with the one angular
momentum, identifying $l=l_1$. Here we consider the two angular
momentum case setting $l_{3}=0$ just for simplicity.

The lower limit $r_{H}$ is given
$r_{H}^{2}=\f{l_{1}^{2}+l_{2}^{2}}{2}
+\s{r_{0}^{4}+\f{(l_{1}^{2}-l_{2}^{2})^{2}}{4}}$, and $f$, $h$,
$\ti{h}$, $\Delta$ and $H_{i} (i=1,2,3)$ are all positive when
$r>r_{H}$. Around the point $\theta=\phi=0$ and $r=r_{H}$, the
metric looks like \eqn\metriceextwo{\eqalign{ds^2\simeq &\f{\beta
r_H^4}{r_0^4\cosh\ap}(r-r_H) d\chi^2+\f{r_0^4\cosh\ap }{\beta
r_H^4}\f{dr^2}{r-r_H}\cr
&+(r_{H}^{2}-l_{1}^{2})\cosh\ap\left[d\theta^2
+\theta^2(d\phi_{1}-\f{l_{1}}{(r_{H}^{2}-l_{1}^{2})\cosh\ap}d\chi)^2\right]\cr
&+(r_{H}^{2}-l_{2}^{2})\cosh\ap\left[d\phi^2
+\phi^{2}(d\phi_{2}-\f{l_{2}}{(r_{H}^{2}-l_{2}^{2})\cosh\ap}d\chi)^2\right],}}
where $\beta=\f{4}{r_H}-\f{2(l_{1}^2+l_{2}^{2})}{r_H^3}$.

The regularity requires the following three identifications
\eqn\periodrotatwo{\eqalign{&(\chi,\phi_{i})\sim
(\chi,\phi_{i}+2\pi),\ \ \ \  (i=1,2) \cr & (\chi,\phi_1,
\phi_2)\sim (\chi+L,\phi_1+2\pi\zeta_{1}, \phi_2+2\pi\zeta_2),}}
where \eqn\lengtwo{ L=\f{2\pi
r_0^4\cosh\ap}{2r_H^3-(l_{1}^2+l_{2}^{2})r_H},\ \ \ \zeta_{i}=
\f{l_{i}r_0^4}{(2r_H^3-(l_{1}^2+l_{2}^{2})r_H)(r_{H}^{2}-l_{i}^{2})}.}
The parameters $\zeta_1, \zeta_2$ take the values within
$0\leq\zeta_1,\zeta_2<1$ and $0\leq \zeta_1+\zeta_2<1$. Suppose that
the twist parameters $\zeta_1, \zeta_2$ take rational values
\eqn\ration{\zeta_i=\f{k_i}{M},\ \ (i=1,2),} where $k_i$ and $M$ are
coprime positive integers. Then the string theory approaches the one
defined on the (generalized) twisted circle $(C^2\times
S^1)/Z_M\times R^{1,4}$ \TaUe \RTF\ in the asymptotic region $r\to
\infty$. Since the point $\zeta_1=\zeta_2=0$ corresponds to the
anti-periodic boundary condition for fermions, the sixteen
supersymmetries are preserved when the following condition is
satisfied \TaUe \RTF \eqn\susysol{\zeta_1+\zeta_2=1.} In the dual
gauge theory side, the Yang-Mills theory becomes $N=2$
supersymmetric. Except these supersymmetric points, the string
theory on the twisted circle includes tachyon field as before \TaUe
\RTF. We wish to claim that in the presence of D3-branes the
endpoint of the closed string tachyon condensation is given by the
twisted AdS bubble solution \rotatetwo.

Let us compute the Casimir energy. The transverse (complex) scalar
in the $C^2$ direction is denoted by $\Phi^i \ (i=1,2)$, and the
other scalars are denoted by $\phi$. Their twisted boundary
conditions are written as \eqn\twisttwo{\Phi^i_{ab}(z+L)=e^{\f{2\pi
i}{M}(a-b+k_i)}\Phi^i_{ab}(z),\ \ \ \ \phi_{ab}(z+L)=e^{\f{2\pi
i}{M}(a-b)}\phi_{ab}(z),} where $0\leq a,b<M$. For the fermions we
similarly find \eqn\twisttwo{\psi^{\pm}_{ab}(z+L)=e^{\f{\pi i(k_1\pm
k_2+M)}{M}}e^{\f{2\pi i(a-b)}{M}}\psi^{\pm}_{ab}(z).}
 When we consider $N$ (fractional) D3-branes with the same value of the
Wilson line, we obtain \eqn\casiketwo{\eqalign{T_{00}^{freeYM}&
=\f{N^2}{\pi^2 L^4}\Biggl[-4\sum_{n=1}^\infty
\f{1}{n^4}-2\sum_{n=1}^\infty \f{\cos\bigl(\f{2\pi
nk_1}{M}\bigr)+\cos\bigl(\f{2\pi nk_2}{M}\bigr)}{n^4}\cr
&+4\sum_{n=1}^\infty\f{\cos\bigl(\f{\pi n(k_1+k_2+M)}{M}\bigr)
+\cos\bigl(\f{\pi n(k_1-k_2+M)}{M}\bigr)}{n^4} \Biggr] \cr
&=-\f{N^2\pi^2}{L^4}
\left[\f{1}{6}-(\zeta_1^2+\zeta_2^2)+\f{4}{3}(\zeta_1^3+ \zeta_2^3)
-\f{1}{2}(\zeta_1^2-\zeta_2^2)^2 \right].}}

On the other hand, in the gravity side, the energy density of this
twisted bubble is given
\eqn\energydensutwo{T_{00}^{gravity}=-\f{\pi^2
r_0^4}{16G^{(10)}_N}=-\f{\pi^2N^2}{8L^4}\cdot
\left(1+\f{(x^2-y^2)^2}{4} \right)^{-2}\cdot
\left(\f{x^2+y^2}{2}+\s{1+\f{(x^2-y^2)^2}{4}}\right)^{-2},} where
$x=l_1/r_0$ and $y=l_2/r_0$. We can again confirm the qualitative
agreement of the behavior of the energy density between the free
Yang-Mills and gravity result. We can check in both sides that
$T_{00}$ vanishes along the supersymmetric points \susysol\ as
expected.

We can also examine the entanglement entropy as in the previous
case. The gravity computation leads to \eqn\gravitymel{\Delta
S_A^{gravity}=-\f{\pi N^2V_1}{(4+(x^2-y^2)^2)L}.} On the other hand,
the free Yang-Mills result reads \eqn\freemel{\Delta
S_A^{freeYM}=-\f{\pi
N^2V_1}{6L}\left(1+2(\zeta_1+\zeta_2)-3(\zeta_1^2+\zeta_2^2)\right).}
We can check the qualitative agreement between them as before. The
limits which approach  backgrounds with sixteen supersymmetries
(i.e. $\zeta_1+\zeta_2=1$ as in \susysol) are given by
\eqn\susylimit{x^2-y^2=\ap^2,\ \ \ x\to\infty, \ \ \ y\to \infty ,}
where $\ap$ is a finite constant. Then the twist parameters are
given by
\eqn\twistpp{\zeta_1=\f{\f{\ap^2}{2}+\s{1+\f{\ap^4}{2}}}{2\s{1+\f{\ap^4}{2}}},\
\ \ \ \zeta_2=\f{-\f{\ap^2}{2}
+\s{1+\f{\ap^4}{2}}}{2\s{1+\f{\ap^4}{2}}}.} In this limit, the free
Yang-Mills result \freemel\ is simplified as follows
\eqn\susyfreeyms{\Delta S_A^{freeYM}\to -\f{\pi
N^2V_1}{L}\cdot\f{1}{4+\ap^4}.} This precisely agrees with the
gravity result \gravitymel. In this comparison, the important point
is that the entropy for the $N=2$ super Yang-Mills is different from
that for the $N=4$ super Yang-Mills. We can also regard this
successful quantitative agreement as a further support for the
assumed holographic calculation \formulaten. It will be a moderate
exercise to extend the above results to the three parameter cases
$l_i\neq 0\ \  (i=1,2,3)$, which include $N=1$ super Yang-Mills
theories.

\subsec{Comparison with Known Results from World-sheet RG-flow
Analysis}

It will also be helpful to compare\foot{We are very grateful to
Takao Suyama for useful discussions about the materials in this
subsection.} the above decay process with the results obtained from
the world-sheet RG-flow \DaGuHeMi \MiTa (for a review see \HeMiTa)
using the gauged linear $\sigma$-model. In this analysis, the decay
of the twisted circle in flat space (i.e. $(R^2\times S^1)/Z_M$) is
considered and thus there are no D3-branes. The vanishing of the
twisted circle $S^1$ is also observed in this world-sheet RG-flow
analysis \MiTa. However, after the circle vanishes, another circle
(called the supersymmetric cycle) $\ti{S}^1$ appears and the
endpoint becomes $R^2\times \ti{S}^1$, where the radius of the new
circle becomes $M$ times that of the twisted circle \DaGuHeMi \CoGu.
This latter process is not included in our AdS counter part \rotate.

Probably, this difference is due to the presence of the cosmological
constant or equally of the D3-branes. Since the conservation of the
twisted sector RR-charges is violated by the closed string tachyon
condensation \MiTa (see also \MaMo\ for the two dimensional orbifold
$C/Z_N$), we expect a large back reaction in the presence of
fractional D3-branes. Indeed, as we have seen, the D-branes which
constitute the AdS twisted bubble \rotate\ are identified with the
fractional D3-branes of the same kind. It will be an interesting
future problem to explore this issue.

\newsec{Null Boundaries in String Theory and Closed String Tachyons}

Up to now, we have discussed static bubble solutions in string
theory. Generally, such a background is described by a complicated
metric and
 RR flux and it is not easy to solve the
corresponding string theory. One way to simplify the
background is to take a particular limit
without ending up with a trivial solution. Consider an
infinite boost of the asymptotically flat bubble solution
 \rotate. Remember that in the previous section
we claimed that
this background \rotate\ is an end point of the closed
 string tachyon condensation. In particular,
we set $l=0$ in \rotate\ for simplicity. This is the
 static bubble solution whose near horizon limit
is the AdS bubble \bubble.

After the infinite boost $t\pm r\to \gamma^{\mp}(t\pm r)$
and $\gamma\to\infty$, we find that
the metric becomes simplified as
 follows \eqn\metricbul{\eqalign{ds^2&\simeq-dt^2+dr^2+d\chi^2+dx_1^2+dx_2^2
+\f{\gamma^2}{4}(t-r)^2d\Omega_5^2\cr
&\to -dt^2+dr^2+d\chi^2+dx_1^2+dx_2^2+(t-r)^2\sum_{i=1}^5 dy_i^2,}}
where the final expression can
be found by noting that the five sphere can be
approximated by $y^i \in R^5$ in the limit $\gamma\to \infty$.
 Furthermore, since the original
radial coordinate is restricted to
the values $r\geq r_0$, the allowed values of $(t,r)$ in
\metricbul\ become after the boost \eqn\allow{r-t>0.}
 Thus this spacetime has a null
boundary\foot{If we treat the light-cone time $t-r$
 as a real time, this background describes a
big crunch-like singularity with $y_i$ compactified
 appropriately, which is almost the same as
a half of the spacetime considered in \LMS.}
 (or light-like boundary) at $r-t=0$.

In general, there has been no systematic
 understanding on what kinds of spacetime
boundaries are allowed in the string theory. Therefore, it will be
helpful to examine various string theory backgrounds with spacetime
boundaries. As we will see below this subject is closely related to
the closed string tachyon condensation since the tachyon wall can be
regarded as a spacetime boundary. At the same time, a non-static
boundary in spacetime offers us a simple time-dependent background
in string theory.

Below we will discuss exactly solvable examples with null boundaries
in critical string theory. They are obtained
 from bulk closed string tachyon condensations\foot{
Clearly, another series of string theory backgrounds with spacetime
boundaries can also be found from orbifold theories.} simpler than
the quasi local tachyon condensations
 relevant for \bubble\ and \rotate. They are
 so simple that their spacetimes except
 the null boundaries are just the ordinary 26 or 10 dimensional
 flat spacetime, where
 the perturbative description of string theory can be done exactly.

\subsec{Null Boundaries in Bosonic String}

Consider the 26 dimensional critical bosonic string. The coordinates
in the bosonic string are denoted by $x_\mu\ \ (\mu=0,1,2,\ddd,25)$
and their world-sheet fields are written as $X_\mu$. We assume the
null (or light-like) linear dilaton\foot{Recently, the null linear
dilaton background in the critical type II string is investigated as
a model of cosmological singularity \CSV.} in this
background\foot{We set $\al=1$ in this paper. The OPE is normalized
such that $X^\mu(z)X^\nu(0)\sim \eta^{\mu\nu}\log z$.}
\eqn\lightdilaton{g_s=e^{-Q(x_0+x_1)}.} Notice that the total
central charge for the world-sheet fields $X_0$ and $X_1$ remains
$c=2$ and thus the background is still critical.

Furthermore we put the Liouville potential
\eqn\liouville{S_L=\mu\int dz^2 e^{-2bX_1},} where $b>0$ is
determined from the relation $Q=b+1/b$. This regulates the strongly
coupled regions at large $X_1$.

Now we perform the infinite Lorentz boost such that the linear
dilaton gradient becomes zero. Explicitly, this is realized by
defining the new boosted coordinates $\ti{x}_0,\ti{x}_1$ as follows
\eqn\boost{\ti{x}_0+\ti{x}_1=\gamma (x_0+x_1),\ \ \ \
\ti{x}_0-\ti{x}_1=\gamma^{-1} (x_0-x_1),} and taking the limit
$\gamma\to \infty$. It is trivial to see $g_s=$const. after this
limit is taken.

After this boost, the Liouville potential looks like \eqn\liaft{
S_L=\mu\int dz^2\ e^{b\gamma(\ti{X}_0-\ti{X}_1)}.} Since we are
taking the limit $\gamma\to \infty$, the potential \liaft\ kills the
half of the spacetime. Thus only the part
\eqn\physical{\ti{x}_1-\ti{x}_0>0,} survives and the fields can
propagate only there. In other words, the closed string tachyon $T$
condenses completely $T\to \infty$ in $\ti{x}_1-\ti{x}_0<0$ and that
part of the spacetime disappears as in Fig.5. On the other hand, the
tachyon field is vanishing for the opposite region
$\ti{x}_1-\ti{x}_0>0$.

\fig{A spacetime with a null boundary $\ti{x}_1-\ti{x}_0=0$ induced
by  a closed string tachyon condensation. The tachyon condenses
completely in the shaded region and the physical spacetime is given
by the opposite half region
$\ti{x}_1-\ti{x}_0>0$.}{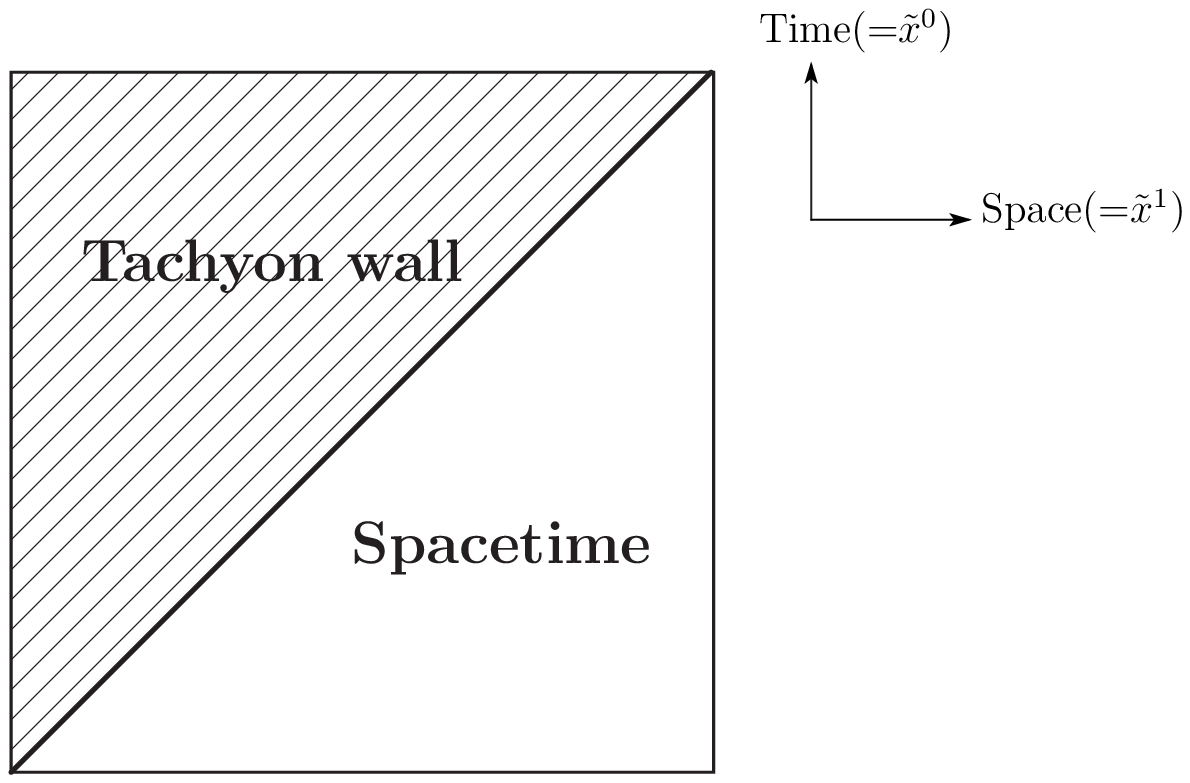}{3truein}

Usually the Liouville potential can be interpreted as a tachyon
wall. The wall in our example after the infinite boost becomes
completely rigid in that the tachyon becomes suddenly infinite when
$\ti{x}_0-\ti{x}_1=0$. The wall moves at the speed of light toward a
static observer (see Fig.5). We can equally obtain the opposite
background (i.e. defined by $\ti{x}_0+\ti{x}_1>0$ instead of
\physical) by flipping the sign of $x_0$.

This background can be regarded as a flat space with a null
boundary. This is because the dilaton and metric is trivial in the
region \physical\ as is clear from the above discussion. This
background will be one of the simplest examples of spacetime
boundaries in the critical bosonic string theory ({\it cf}.
analogous models \KaSt \GTT \DDLM \TaLD\ in 2D string theory). These
arguments can be easily generalized to the critical type 0 string
theory where a similar type of closed string tachyon field exists.

\subsec{Null Boundaries in Type II String}

It is more interesting to ask if a similar null boundary in the flat
space is allowed in the critical type II superstring. In this case
we need to take an additional coordinate $X_2$ into account. In
order to obtain the Liouville potential, we compactify $X_2$ such
that we can put the $N=2$ Liouville potential ($\Phi=X_1+iX_2$)
\eqn\liount{S_L=\mu\int dz^2 d\theta^2 e^{-\f{\Phi}{Q}}+(h.c.),} in
the null linear dilaton background \lightdilaton. By boosting as
before, we again find that only the half of spacetime \physical\
survives. Since we can choose $Q$ independently, we can decompactify
the circle\foot{Remember the radius of circle is proportional to
$Q$.}. Therefore we can conclude that we can put a null boundary
also in type II string theory as in Fig.5.

We can obtain the same result by taking T-dual in the circle
direction $X_2$. The FZZ duality leads to the equivalent background
with the following non-trivial metric and dilaton \HoKa
\eqn\metric{ds^2=-(dx_0)^2+(dx_1)^2+\f{1}{Q^2}\tanh^2
(Qx_1)(d\theta)^2,\ \ \ \ \ \ \ \ g_s=\f{e^{Qx_0}}{\cosh(Qx_1)},}
where $\theta$ is compactified such that $\theta\sim \theta+2\pi$.
Also the value of $x_1$ is restricted as $x_1\geq 0$.

Now we perform the previous boost \boost.
When $\gamma$ is very large, $x_{0}\sim x_{1}$ becomes
large too, and thus we can approximate the string coupling as
$g_s\sim
e^{-Q(x_0+x_1)}=e^{-Q\gamma^{-1}(\ti{x}_0+\ti{x}_1)}.$ The metric is
also approximated by \eqn\metriclimi{ ds^2\simeq
-(d\ti{x}_0)^2+(d\ti{x}_1)^2
+\f{1}{Q^2}\tanh^2\left(\f{Q\gamma}{2}(\ti{x}_1-\ti{x}_0)\right)(d\theta)^2.}
The restriction $x_1\geq 0$ means \eqn\restm{\ti{x}_1-\ti{x}_0>0.}
Finally we take the limit $\gamma\to \infty$.
Then this spacetime is identified with
the flat space with a rigid wall (or boundary) at
$\ti{x}_1-\ti{x}_0=0$, as expected. Notice that
this argument is analogous to our
previous one \metricbul.

 After the infinite boost, we recover
the flat type II string in the half spacetime \restm\ and thus all
of the 32 supersymmetries are preserved in the bulk points\foot{Note
also that the dilaton is constant after the boost. In the flat
background with the null linear dilaton,
 only 16 supersymmetries are
preserved \CSV.}. However, the supersymmetry is completely broken at
the rigid wall $\ti{x}_0+\ti{x}_1=0$.

\subsec{More Null Boundaries}

In the above examples, the metric, dilaton and
tachyon are the same as those in the ordinary
flat spacetime except the boundaries. Thus we
expect that the equation of motion should be
 satisfied even if we put multiple boundaries
 in the flat spacetime. For example, we
can construct a vacuum restricted to the region
\eqn\belt{a<\ti{x}_1-\ti{x}_0<b.}
This represents a spacetime where
a finite interval moving at the
speed of light as in the left figure
of Fig.6.

An explicit construction of this spacetime is to start with the
space-like Liouville term \liouville\ as well as the time-like
Liouville term \eqn\timeli{\nu \int dz^2\ e^{-2\beta X^0},} where
$Q=\beta-1/\beta$. After boost we obtain $T_{closed}\sim \mu
e^{b\gamma(\ti{x}_0-\ti{x}_1)} +\nu
e^{\beta\gamma(\ti{x}_1-\ti{x}_0)}$. Thus if we assume an
appropriate limit of $\mu/\nu$ we indeed find the restriction \belt.

A more ambitious example may be the spacetime with two different
types of the null boundaries i.e. $\ti{x}^+>0$ and $\ti{x}^->0$ at
the same time (see the right figure of Fig.6). Except the point
$\ti{x}^+=\ti{x}^-=0$, the equation of motion is clearly satisfied
as in the previous argument. However, since near the origin the
boundaries coincide with each other, there is a possibility that we
have to modify the solution to take stringy backreactions into
account. If we neglect this issue, it is clear that this background
describes a universe created at $t=x=0$ and expands at the speed of
light. This construction of the spacetime via the closed string
tachyon condensation suggests the recently advertised idea that a
spacetime is an emergent object. We leave further studies of this
background for a future problem.

\fig{Examples of spacetimes with two null boundaries induced by
closed string tachyon condensation. The two tachyon walls move in
the same (or opposite) direction in the left (or right)
spacetime.}{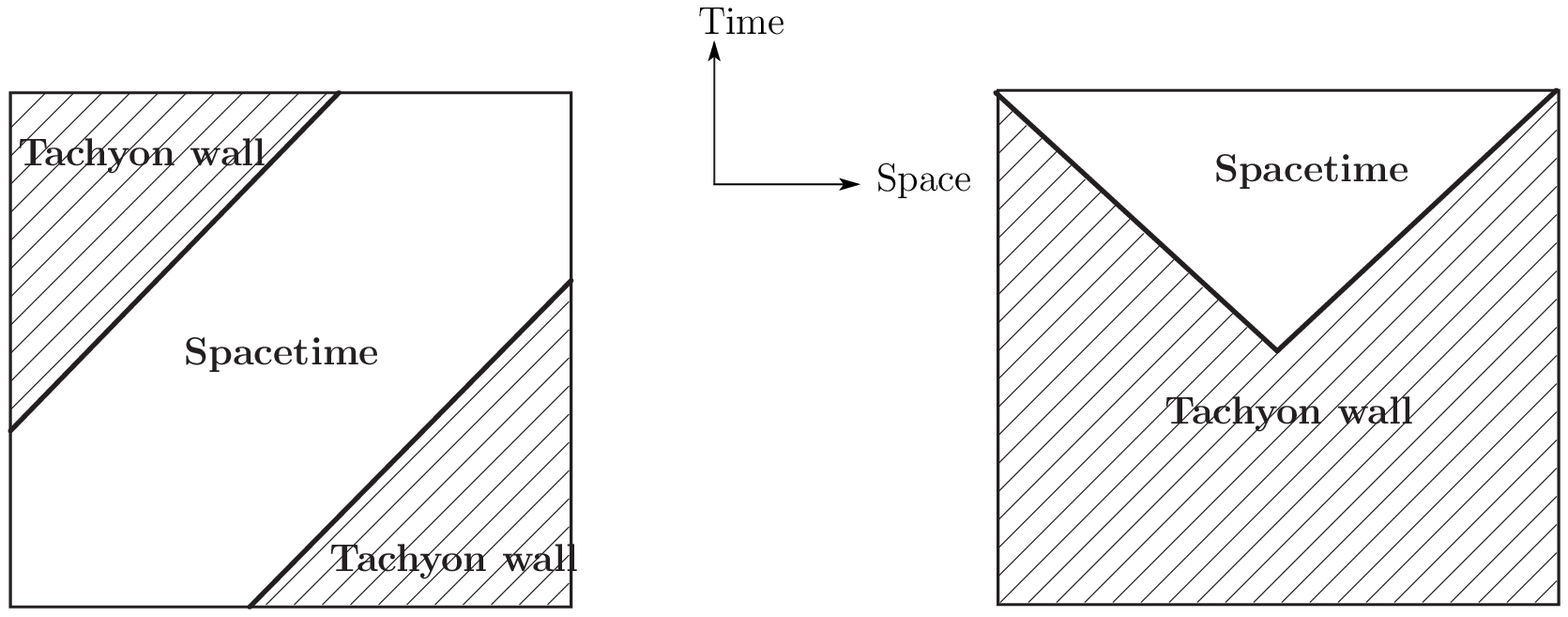}{5truein}

\subsec{Effective Action Argument of Null Boundaries}

It is also useful to see how the spacetime with the null boundary
makes sense from the viewpoint of the effective low energy gravity
theory with a closed string tachyon. Remember the usual caveat that
we cannot trust this analysis quantitatively since the tachyon mass
is of order string scale and that, strictly speaking, we have to
take higher derivative terms into account.

We assume the following model as a candidate of an effective action
for the $26D$ bosonic string \eqn\yzaction{S=\f{1}{2\kappa^2}\int
(dx)^{26} \s{-G} e^{-2\Phi} \left(R+4(\de_\mu \Phi)^2-f(T)(\de_\mu
T)^2-2V(T)\right).} If we consider the particular case $f(T)=1$,
this is exactly the same as the one \YaZw\ considered by Yang and
Zwiebach (see also \Ts \SuRG). We extended this model to allow any
function $f(T)$ with the requirement $f(0)=1$. This is because
physically we may be interested in the possibility $f(\infty)=0$,
which is motivated from the speculation that the complete tachyon
condensation $T=\infty$ annihilates the spacetime and that there
should be no degree of freedom as in the open string case \Se.

The tachyon potential is supposed to satisfy the following
properties \YaZw\ \eqn\property{V(0)=V'(0)=0,} and
\eqn\propertytwo{V(\infty)=V'(\infty)=0.} The first conditions in
\property\ and \propertytwo\ are required in order to satisfy the
dilaton equation of motion. The vacuum $T=0$ represents the ordinary
(tachyonic) closed string vacuum, while another one $T=\infty$
represents the one after the tachyon condensation. The second
conditions in \property\ and \propertytwo\ assure the absence of
tad-pole of the tachyon field $T$. For example, the potential
 $V(T)=T^2 e^{-T}$ \Ts\ proposed by Tseytlin indeed satisfies all of
 these conditions. However, it is fair to say that one of the assumptions
$V(\infty)=0$ has not been completely well-established (some counter
evidences have been recently
 discussed in \MoYa). Even if $V(\infty)=0$ is not correct,
 we believe that a certain (non-substantial) modification of our
analysis below can be done to show that the null boundary is an
allowed solution.

The equation of motions are given by \eqn\eoms{\eqalign{ &R_{\mu\nu}
+2\nabla_{\mu}\nabla_{\nu}\Phi-f(T)\de_{\mu}T \de_{\nu}T=0, \cr &
2f(T)\nabla_{\mu}\de^\mu
T+f'(T)(\de_{\mu}T)^2-4f(T)\de_\mu\Phi\de^\mu T-2V'(T)=0, \cr &
\nabla^2 \Phi-2(\de_{\mu}\Phi)^2-V(T)=0.}}

Now we would like to confirm that the null boundary background
indeed satisfies the equation of motions \eoms . Suppose that the
metric is flat $G_{\mu\nu}=\eta_{\mu\nu}$ and the dilaton and
tachyon only depends on the light-cone coordinate $x^+$
\eqn\lightde{\Phi=\Phi(x^+),\ \ \ T=T(x^+).} Then the first equation
in \eoms\ leads to \eqn\diffeo{2\de_{+}^2\Phi=f(T)(\de_+ T)^2.} We
can choose the tachyon field of the following form
\eqn\formta{T=\mu\ e^{\lambda x^+},} which is the same as in the
boosted Liouville model \liaft. The second and third equation in
\eoms\ are equivalent to \eqn\eom{V(T)=V'(T)=0.} Therefore if we
take the limit $\lambda\to \infty$, then the tachyon profile
\formta\ satisfies\foot{ In this limit the dilaton becomes trivial
$\Phi(x^+)\to 0$ if we assume the profile $f(T)\sim e^{-T}$.} \eom\
because of the properties \property\ and \propertytwo.

We would also like to note that we can add the null dilaton
$\Phi(x^+)_{null}=Qx^+$ for any $Q$ with the equation of motions
satisfied. In other word, we can start with the null dilaton
background and cut off the strongly coupled region by putting the
null boundary.

It can be easily checked that the above argument is also true for
the model obtained by Tseytlin (setting $D=26$) via the sigma-model
approach \Ts\ \eqn\actionte{S=\int dx^{D}\s{g}\left(
V(T)e^{\f{4\Phi}{D-2}}+\al F(T)(\de_{\mu}
T)^2-\f{\al}{2}[R-\f{4}{D-2}(\de_{\mu} \Phi)^2]\right),} where $V$
and $F$ are explicitly given by \eqn\vfex{V(T)=-2T^2 e^{-2T},\ \ \ \
F(T)=2(1-T)e^{-2T}.}

Finally we would like to comment on the total energy of the
spacetime. As far as we assume \propertytwo\ as well as \property,
it is clear that the total energy is divergent as the kinetic energy
is infinite due to the step function-like behavior of the tachyon
field $T$. This is essentially because we boosted the tachyon wall
infinitely and this will not be a serious problem as far as we start
with this background from the beginning. The background cannot be
created dynamically from the ordinary vacuum of flat spacetime.
However, if the assumption $V(\infty)=0$ in \propertytwo\ is not
correct and the tachyon vacuum has a negative energy, there is a
possibility that the total energy is finite and even negative.
Naively, our physical intuition influenced by the present knowledge
of the open string tachyon condensation \Se\ suggests the second
possibility i.e. $V(\infty)<0$. However, this speculation may be
wrong in the presence of dilaton as discussed in \YaZw\ and its
final answer will be an important open problem.

\subsec{D-brane Analogue}

As far as we consider an unstable D-brane (i.e. a D-brane with an
open string tachyon) in type II or other string theories, we can
apply the same boost argument in the presence of the null dilaton
and the boundary Liouville term $\mu_B \int_{\de\Sigma} dz\
e^{-bX_1}$ \FaZaZa. Then we can construct a D-brane with a null
boundary in the same way.

\newsec{Conclusions and Discussions}

More than half of this paper has been devoted to explore evidences
for the conjectured scenario that unstable near horizon geometries
of D-branes may decay into stable AdS bubbles with the same
asymptotic geometry via the closed string tachyon condensation. In
particular we examined the novel quantity called the entanglement
entropy in both gravity and Yang-Mills side, which measures the
degree of freedom. We show that the entropy decreases under the
tachyon condensations in explicit examples as we expect. A new
example discussed in this paper is the twisted AdS bubble obtained
by the double Wick rotation of the rotating black 3-brane solution.
This string theory background includes a closed string tachyon
localized both in the IR region and in the north pole of the $S^5$.
This tachyon is very similar to the one found in the twisted circle
(or Melvin background). We can also say that the tachyon
condensation process from the AdS twisted circle to the AdS twisted
bubble is dual to the shift of the Wilson line expectation values
from those for the bulk D3-branes to those for the fractional
D3-branes if $k+M$ is even.

It is known that the AdS bubble has a lower energy than the one of
the pure $AdS_5$ when we impose the anti-periodic boundary condition
for fermions. The Casimir energy of the free $N=4$ super Yang-Mills
on $S^1\times R^3$ with the same boundary condition for fermions
agrees with the energy of AdS bubble in the gravity computation up
to the factor of $\f{4}{3}$. In the near extremal limit of the
twisted AdS bubble example we find that this ratio is given by
$\f{9}{8}$ and thus becomes much more closer to $1$.

We obtain a similar behavior also for the entanglement entropy. The
free Yang-Mills/gravity ratio of the entropy becomes $\f{2}{3}$ for
the AdS bubble. Remarkably, in the near extremal limit the ratio
becomes precisely $1$. All of these results indicate that the
(twisted) AdS bubble is the true gravity dual geometry
corresponding
to the Yang-Mills theory on $S^1\times R^3$ with the twisted
boundary conditions.

In the final part of this paper we have discussed null spacetime
boundaries in string theory. We observe that the tachyon walls or
bubbles in a null linear dilaton background lead to such null
boundaries after an infinite boost. They are exactly solvable
time-dependent backgrounds since they are described by the Liouville
theory before we take the boost. Because the metric of this
spacetime is strictly flat except the sharp tachyon wall, it is
natural to expect a direct string theoretic description of this
spacetime without using the Liouville theory. In the light-cone
gauge it may be described by the restriction $\tau>0$ of the
world-sheet time $\tau(=X^+)$. The covariant string description
remains as a future problem.

\vskip .5in

\centerline{\bf Acknowledgments} We are grateful to T. Hirata, T.
Muto, S. Ryu, A. Shirasaka, N. Tanahashi and S. Terashima for
helpful discussions, and especially to Y. Hikida and T. Suyama for
important comments. We also thank T. Harmark very much for useful
correspondence. We are very grateful to G. Horowitz for pointing an
important error in eq.(2.16) in our first version of this paper.

The work of TT is supported in part by JSPS Grant-in-Aid for
Scientific Research No.18840027.

\vskip .3in

\listrefs

\end